\documentclass[aps,physrev,twocolumn,notitlepage,nofootinbib,superscriptaddress,longbibliography]{revtex4-2}

\usepackage[dvipsnames]{xcolor}
\usepackage{framed}
\definecolor{shadecolor}{rgb}{0.9,0.9,0.9}

\usepackage{mathtools}
\usepackage{amsmath}
\usepackage[shortlabels]{enumitem}

\usepackage{graphicx,epic,eepic,epsfig,amsmath,latexsym,amssymb,verbatim,color}
 
\usepackage{amsfonts}       
\usepackage{nicefrac}       

\usepackage{amsmath}
\usepackage{bbm}

\usepackage{float}
\usepackage{tikz}
\usetikzlibrary{chains}
\usetikzlibrary{fit}
\usepackage{pgflibraryarrows}		
\usepackage{pgflibrarysnakes}		

\usepackage{epsfig}
\usetikzlibrary{shapes.symbols,patterns} 
\usepackage{pgfplots}

\usepackage[strict]{changepage}
\usepackage{hyperref}
\hypersetup{colorlinks=true,citecolor=blue,linkcolor=blue,filecolor=blue,urlcolor=blue,breaklinks=true}

\usepackage[marginal]{footmisc}
\usepackage{url}
\usepackage{theorem}

\newtheorem{proposition}{Proposition}

\newtheorem{theorem}[proposition]{Theorem}

\newtheorem{corollary}[proposition]{Corollary}

\def\squareforqed{\hbox{\rlap{$\sqcap$}$\sqcup$}}
\def\qed{\ifmmode\squareforqed\else{\unskip\nobreak\hfil
\penalty50\hskip1em\null\nobreak\hfil\squareforqed
\parfillskip=0pt\finalhyphendemerits=0\endgraf}\fi}
\def\endenv{\ifmmode\;\else{\unskip\nobreak\hfil
\penalty50\hskip1em\null\nobreak\hfil\;
\parfillskip=0pt\finalhyphendemerits=0\endgraf}\fi}
\newenvironment{proof}{\noindent \textbf{{Proof~} }}{\hfill $\blacksquare$}

\newcounter{remark}

\newcounter{example}

\mathchardef\ordinarycolon\mathcode`\:
\mathcode`\:=\string"8000
\def\vcentcolon{\mathrel{\mathop\ordinarycolon}}
\begingroup \catcode`\:=\active
  \lowercase{\endgroup
  \let :\vcentcolon
  }

\usepackage{cleveref}
\usepackage{graphicx}
\usepackage{xcolor}

\RequirePackage[framemethod=default]{mdframed}
\newmdenv[skipabove=7pt,
skipbelow=7pt,
backgroundcolor=darkblue!15,
innerleftmargin=5pt,
innerrightmargin=5pt,
innertopmargin=5pt,
leftmargin=0cm,
rightmargin=0cm,
innerbottommargin=5pt,
linewidth=1pt]{tBox}

\newmdenv[skipabove=7pt,
skipbelow=7pt,
backgroundcolor=red!15,
innerleftmargin=5pt,
innerrightmargin=5pt,
innertopmargin=5pt,
leftmargin=0cm,
rightmargin=0cm,
innerbottommargin=5pt,
linewidth=1pt]{rBox}

\newmdenv[skipabove=7pt,
skipbelow=7pt,
backgroundcolor=blue2!25,
innerleftmargin=5pt,
innerrightmargin=5pt,
innertopmargin=5pt,
leftmargin=0cm,
rightmargin=0cm,
innerbottommargin=5pt,
linewidth=1pt]{dBox}
\newmdenv[skipabove=7pt,
skipbelow=7pt,
backgroundcolor=darkkblue!15,
innerleftmargin=5pt,
innerrightmargin=5pt,
innertopmargin=5pt,
leftmargin=0cm,
rightmargin=0cm,
innerbottommargin=5pt,
linewidth=1pt]{sBox}
\definecolor{darkblue}{RGB}{0,76,156}
\definecolor{darkkblue}{RGB}{0,0,153}
\definecolor{blue2}{RGB}{102,178,255}
\definecolor{darkred}{RGB}{195,0,0}

\newcommand{\nc}{\newcommand}
\nc{\rnc}{\renewcommand}
\nc{\lbar}[1]{\overline{#1}}
\nc{\bra}[1]{\langle#1|}
\nc{\ket}[1]{|#1\rangle}
\nc{\ketbra}[2]{|#1\rangle\!\langle#2|}
\nc{\braket}[2]{\langle#1|#2\rangle}

\nc{\proj}[1]{| #1\rangle\!\langle #1 |}
\nc{\avg}[1]{\langle#1\rangle}
\nc{\smfrac}[2]{\mbox{$\frac{#1}{#2}$}}
\nc{\tr}{\operatorname{Tr}}
\nc{\ox}{\otimes}
\nc{\dg}{\dagger}
\nc{\dn}{\downarrow}
\nc{\cA}{{\cal A}}
\nc{\cB}{{\cal B}}
\nc{\cC}{{\cal C}}
\nc{\cD}{{\cal D}}
\nc{\cE}{{\cal E}}
\nc{\cF}{{\cal F}}
\nc{\cG}{{\cal G}}
\nc{\cH}{{\cal H}}
\nc{\cI}{{\cal I}}
\nc{\cJ}{{\cal J}}
\nc{\cK}{{\cal K}}
\nc{\cL}{{\cal L}}
\nc{\cM}{{\cal M}}
\nc{\cN}{{\cal N}}
\nc{\cO}{{\cal O}}
\nc{\cP}{{\cal P}}
\nc{\cQ}{{\cal Q}}
\nc{\cR}{{\cal R}}
\nc{\cS}{{\cal S}}
\nc{\cT}{{\cal T}}
\nc{\cU}{{\cal U}}
\nc{\cV}{{\cal V}}
\nc{\cX}{{\cal X}}
\nc{\cY}{{\cal Y}}
\nc{\cZ}{{\cal Z}}
\nc{\cW}{{\cal W}}
\nc{\csupp}{{\operatorname{csupp}}}
\nc{\qsupp}{{\operatorname{qsupp}}}
\nc{\var}{{\operatorname{var}}}
\nc{\rar}{\rightarrow}
\nc{\lrar}{\longrightarrow}
\nc{\polylog}{{\operatorname{polylog}}}
\nc{\wt}{{\operatorname{wt}}}
\nc{\av}[1]{{\left\langle {#1} \right\rangle}}
\nc{\supp}{{\operatorname{supp}}}

\nc{\argmin}{{\operatorname{argmin}}}

\def\x{\xi}

\nc{\RR}{{{\mathbb R}}}
\nc{\CC}{{{\mathbb C}}}
\nc{\FF}{{{\mathbb F}}}
\nc{\NN}{{{\mathbb N}}}
\nc{\ZZ}{{{\mathbb Z}}}
\nc{\PP}{{{\mathbb P}}}
\nc{\QQ}{{{\mathbb Q}}}
\nc{\UU}{{{\mathbb U}}}
\nc{\EE}{{{\mathbb E}}}
\nc{\id}{{\operatorname{id}}}

\nc{\CHSH}{{\operatorname{CHSH}}}

\nc{\be}{\begin{equation}}
\nc{\ee}{{\end{equation}}}
\nc{\bea}{\begin{eqnarray}}
\nc{\eea}{\end{eqnarray}}
\nc{\<}{\langle}
\rnc{\>}{\rangle}
\nc{\rU}{\mbox{U}}

\nc{\ob}[1]{#1}

\nc{\SEP}{{\text{\rm SEP}}}
\nc{\NS}{{\text{\rm NS}}}
\nc{\LOCC}{{\text{\rm LOCC}}}
\nc{\PPT}{{\text{\rm PPT}}}
\nc{\EXT}{{\text{\rm EXT}}}
\nc{\Sym}{{\operatorname{Sym}}}

\nc{\ERLO}{{E_{\text{r,LO}}}}
\nc{\ERLOCC}{{E_{\text{r,LOCC}}}}
\nc{\ERPPT}{{E_{\text{r,PPT}}}}
\nc{\ERLOCCinfty}{{E^{\infty}_{\text{r,LOCC}}}}
\nc{\Aram}{{\operatorname{\sf A}}}

\usepackage{tikz}
\usepackage{hyperref}
\hypersetup{colorlinks=true,citecolor=blue,linkcolor=blue,filecolor=blue,urlcolor=blue,breaklinks=true}

\makeatletter
\def\grd@save@target#1{%
  \def\grd@target{#1}}
\def\grd@save@start#1{%
  \def\grd@start{#1}}
\tikzset{
  grid with coordinates/.style={
    to path={%
      \pgfextra{%
        \edef\grd@@target{(\tikztotarget)}%
        \tikz@scan@one@point\grd@save@target\grd@@target\relax
        \edef\grd@@start{(\tikztostart)}%
        \tikz@scan@one@point\grd@save@start\grd@@start\relax
        \draw[minor help lines,magenta] (\tikztostart) grid (\tikztotarget);
        \draw[major help lines] (\tikztostart) grid (\tikztotarget);
        \grd@start
        \pgfmathsetmacro{\grd@xa}{\the\pgf@x/1cm}
        \pgfmathsetmacro{\grd@ya}{\the\pgf@y/1cm}
        \grd@target
        \pgfmathsetmacro{\grd@xb}{\the\pgf@x/1cm}
        \pgfmathsetmacro{\grd@yb}{\the\pgf@y/1cm}
        \pgfmathsetmacro{\grd@xc}{\grd@xa + \pgfkeysvalueof{/tikz/grid with coordinates/major step}}
        \pgfmathsetmacro{\grd@yc}{\grd@ya + \pgfkeysvalueof{/tikz/grid with coordinates/major step}}
        \foreach \x in {\grd@xa,\grd@xc,...,\grd@xb}
        \node[anchor=north] at (\x,\grd@ya) {\pgfmathprintnumber{\x}};
        \foreach \y in {\grd@ya,\grd@yc,...,\grd@yb}
        \node[anchor=east] at (\grd@xa,\y) {\pgfmathprintnumber{\y}};
      }
    }
  },
  minor help lines/.style={
    help lines,
    step=\pgfkeysvalueof{/tikz/grid with coordinates/minor step}
  },
  major help lines/.style={
    help lines,
    line width=\pgfkeysvalueof{/tikz/grid with coordinates/major line width},
    step=\pgfkeysvalueof{/tikz/grid with coordinates/major step}
  },
  grid with coordinates/.cd,
  minor step/.initial=.2,
  major step/.initial=1,
  major line width/.initial=2pt,
}
\makeatother

\usepackage{thmtools}
\usepackage{thm-restate}
\usepackage{etoolbox}
\makeatletter
\def\problem@s{}
\newcounter{problems@cnt}

\newcommand{\allproblems}{\problem@s}
\makeatother

\usepackage{amsmath,amsfonts,amssymb,array,graphicx,mathtools,multirow,bm,times,tcolorbox,relsize,booktabs, rotating, cprotect}
\usepackage[utf8]{inputenc}
\usepackage[T1]{fontenc}
\usepackage{tikz}
\usetikzlibrary{quantikz2}

\usepackage[export]{adjustbox}
\usepackage[ruled, vlined, linesnumbered]{algorithm2e}

\definecolor{colortwo}{rgb}{0.4,0.77,0.17}
\definecolor{colorthree}{rgb}{0.01,0.51,0.93}

\allowdisplaybreaks

\begin{document}
\title{Near-Optimal Simultaneous Estimation of Quantum State Moments}

\author{Xiao Shi}\thanks{X.S. and J.J. contributed equally to this work.}
\author{Jiyu Jiang}
\thanks{X.S. and J.J. contributed equally to this work.}
\author{Xian Wu}
\author{Jingu Xie}
\author{Hongshun Yao}
\email{yaohongshun2021@gmail.com}
\author{Xin Wang}
\email{felixxinwang@hkust-gz.edu.cn}
\affiliation{Thrust of Artificial Intelligence, Information Hub, The Hong Kong University of Science and Technology (Guangzhou), Guangzhou 511453, China}




\date{\today}

\begin{abstract}
Estimating nonlinear properties such as Rényi entropies and observable-weighted moments serves as a central strategy for spectrum spectroscopy, which is fundamental to property prediction and analysis in quantum information science, statistical mechanics, and many-body physics. However, existing approaches are susceptible to noise and require significant resources, making them challenging for near-term quantum hardware. In this work, we introduce a framework for resource-efficient simultaneous estimation of quantum state moments via qubit reuse. For an $m$-qubit quantum state $\rho$, our method achieves the simultaneous estimation of the full hierarchy of moments $\text{Tr}(\rho^2), \dots, \text{Tr}(\rho^k)$, as well as arbitrary polynomial functionals and their observable-weighted counterparts. By leveraging qubit reset operations, our core circuit for simultaneous moment estimation requires only $2m+1$ physical qubits and $\mathcal{O}(k)$ CSWAP gates, achieving a near-optimal sample complexity of $\mathcal{O}(k \log k / \varepsilon^2)$. We demonstrate this protocol's utility by showing that the estimated moments yield tight bounds on a state's maximum eigenvalue and present applications in quantum virtual cooling to access low-energy states of the Heisenberg model. Furthermore, we show the protocol's viability on near-term quantum hardware by experimentally measuring higher-order Rényi entropy on a superconducting quantum processor. Our method provides a scalable and resource-efficient route to quantum system characterization and spectroscopy on near-term quantum hardware.
\end{abstract}

\maketitle


\newpage

\section{Introduction}
Precise estimation of polynomial functionals of an $m$-qubit state $\rho$, most notably the purity $\operatorname{Tr}(\rho^{2})$ and its higher-order moments $\operatorname{Tr}(\rho^{k})$ that determine Rényi entropies of order $k$, is a cornerstone of quantum information science~\cite{Artur2002Direct,van2012measuring,yirka2021qubit}, statistical mechanics~\cite{Lee1959many} and many-body physics~\cite{Abanin2019colloquium}.
These moments underpin standard quantifiers of entanglement and coherence~\cite{Streltsov2015measuring,pathania2022quantifying,zhang2024quantification} and act as primitives in randomness certification protocols~\cite{buhrman2001quantum,islam2015measuring,Johri2017Entanglement}.  
More generally, any real polynomial  $f(\rho)=\sum_{j=1}^{k}\alpha_j\operatorname{Tr}\bigl(\rho^{j}\bigr)$ inherits the same practical importance: fast access to $f(\rho)$ enables, for example, the evaluation of integer Rényi or Tallis entropies~\cite{luitz2014Improving,Sathyawageeswar2021quantum,Wang2024newquantum,liu2025estimating}, spectrum reconstruction~\cite{luitz2014Improving,Donnell2016efficient}, entanglement spectroscopy~\cite{islam2015measuring,Johri2017Entanglement,kokail2021entanglement}, and the estimation of other nonlinear functionals~\cite{yao2024nonlinear,chen2025simultaneous}. A still richer family of quantities arises when the moments are ``dressed’’ with an observable $O$, giving the hybrid moments $\operatorname{Tr}(O\rho^{k})$.  
These objects lie at the core of Gibbs-state preparation~\cite{Wang2021Variational,Consiglio2024variational}, quantum virtual cooling and distillation protocols~\cite{Cotler2019Quantum,william2021Virtual}, and have promising implications for quantum machine learning, chemistry, and cryptography~\cite{Maria2015introduction,Huang2021Information,cao2019quantum,weidman2024quantum,Nicolas2002quantum,shenoy2017quantum}.  
Developing hardware-friendly schemes that can accurately evaluate both $\operatorname{Tr}(\rho^{k})$ and its observable-weighted counterpart, therefore, constitutes a pressing goal for near-term quantum technologies.

A variety of strategies have been explored for probing nonlinear properties of quantum states.
Quantum state tomography~\cite{paris2004quantum,cramer2010efficient} reconstructs the full density matrix and therefore gives access to any desired functional while serving as a benchmark for device calibration. 
Despite advances based on compressed sensing~\cite{Gross2010quantum, Ganguli2010Statistical}, neural-network post-processing~\cite{Torlai2020precise, schmale2022efficient}, and other heuristics tailored for structured states like tensor networks~\cite{10806706,guo2024quantum}, its measurement settings and classical workload still grow exponentially with the number of qubits for generic states~\cite{Donnell2016efficient,haah2016sample,torlai2018neural,chen2023when}.
An opposite philosophy is adopted by the generalised SWAP test, which extracts moments through a single cyclic-permutation measurement and thus keeps sample complexity low~\cite{Artur2002Direct,bruni2004measuring}.  The price is hardware overhead: one needs $k$ mutually coherent copies of the state, inflating either circuit width ($km$ qubits) or coherence time.  Recent experiments, however, demonstrate that mid-circuit qubit reset and routing can recycle qubits and make SWAP-style primitives viable on superconducting and trapped-ion platforms~\cite{Magnard2018fast,decross2023qubit}.  
A third line of work, classical shadow tomography, replaces deep circuits by shallow random Clifford layers and estimates up to $M$ observables with merely ${\cal O}(\log M/\varepsilon^{2})$ copies~\cite{huang2020predicting}.  Shallow-shadow and noise-robust variants have further reduced depth and gate sensitivity~\cite{koh2022classical,Christian2024shallow}, yet the estimator variance can still blow up with the moment order $k$ or the operator norm $\|O\|$, limiting accuracy for high-order functionals. 
While recent advances in derandomization, symmetry utilization, and estimator optimization have sought to mitigate this issue for specific tasks~\cite{huang2021efficient,caprotti2024optimizing}, it remains a fundamental bottleneck for generic high-order moment estimation.

Building on these insights, Ref.~\cite{chen2025simultaneous} took an important conceptual step by proving that there exists a family of pairwise commuting observables whose joint measurement yields in one shot the whole set of hybrid moments $\{\operatorname{Tr}(O\rho^{i})\}_{i=1}^{k}$ and, more generally, an entire collection of polynomial functionals $\{f_{i}(O,\rho)\}_{i=1}^{n_f}$.  
The authors first establish an information-theoretic lower bound: for any observable $O$, estimating a single moment $\operatorname{Tr}(O\rho^{k})$ to additive error $\varepsilon$ requires $\Omega\bigl(k\|O\|^{2}/\varepsilon^{2}\bigr)$ copies of $\rho$.  
They then show that the same commuting-measurement scheme can simultaneously estimate the full list $\operatorname{Tr}(O\rho),\operatorname{Tr}(O\rho^{2}),\dots,\operatorname{Tr}(O\rho^{k})$ with accuracy $\varepsilon$ using only ${\cal O}\bigl(k\log k\|O\|^{2}/\varepsilon^{2}\bigr)$ copies, just a logarithmic factor above the single-moment lower bound, which implies that multi-moment estimation is nearly as easy as estimating one moment in isolation.  
Although this achieves near-optimal sample complexity, the study does not provide an explicit circuit implementation, so essential hardware metrics such as qubit count, depth, and connectivity remain unspecified and may, in practice, nullify the promised efficiency.

In this work, we advance practical quantum property spectroscopy and system characterization by introducing an efficient quantum circuit designed for the simultaneous estimation of multiple moments and general nonlinear functionals of a quantum state $\rho$, and presenting an experimental demonstration on superconducting quantum computers.

By leveraging qubit reset techniques~\cite{decross2023qubit,liu2025optimally}, our approach maintains a remarkably low ancilla qubit overhead while achieving near-optimal sample complexity. Our main results for a given $m$-qubit state $\rho$ are summarized below:
\begin{itemize}
\item For the simultaneous estimation of the first $k$ moments $\{\operatorname{Tr}(\rho^{i})\}_{i=1}^{k}$, we construct a circuit using only two state registers ($2m$ qubits total) plus a single ancilla qubit. 
Crucially, the circuit width is independent of the moment order $k$, a feat enabled by qubit reset that circumvents the prohibitive $\mathcal{O}(km)$ qubit requirement of naive multi-copy schemes. 
This resource efficiency is achieved with a circuit depth of $\mathcal{O}(k)$ while retaining the near-optimal sample complexity of $\mathcal{O}\bigl(k\log k/\varepsilon^{2}\bigr)$.

\item We generalize this protocol to directly estimate any $k$-th order polynomial functional $f(\rho)$ via developing the CSWAP-based data reuploading technique. This powerful generalization maintains the constant $2m$ qubit cost for state registers, requiring only a modest $\mathcal{O}(\log k)$ ancilla qubits, a feat also made possible by qubit reset. The circuit depth remains $\mathcal{O}(k)$, and the sample complexity is $\mathcal{O}\bigl(k\|f\|_1^2/\varepsilon^{2}\bigr)$. Furthermore, we show how to estimate $n_f$ such functionals simultaneously with only a minor $\log(\min\{k, n_f\})$ factor increase in samples.

\item Finally, we extend our protocols to estimate observable-weighted functionals, such as $f_i(O,\rho)=\sum_j\alpha_{ij}\mathrm{Tr}\bigl(O\rho^{j}\bigr)$. 
We introduce two distinct strategies, offering a powerful trade-off between circuit complexity and sample efficiency. 
The first approach is particularly straightforward to implement, as it requires zero additional circuit resources; the observable is incorporated simply by performing measurements on additional qubits at the end of the original circuit. This hardware frugality, however, comes at the cost of a higher sample complexity, which scales quadratically with $\|O\|_{\ell_1(\mathcal{P})}$. This quantity, which we term the Pauli-weighted $\ell_1$-norm, is calculated by first decomposing the observable $O$ into a sum of Pauli strings and then summing the absolute values of all the coefficients $\alpha_p$. 
In contrast, a second, more sample-efficient approach achieves the optimal scaling with the spectral norm, $\|O\|^2$, by introducing a minimal hardware overhead of just one additional ancilla qubit and two controlled-observable operations. The full sample complexity for estimating $n_f$ functions simultaneously with this latter scheme is $\mathcal{O}\Bigl(k C_O\max\limits_{1\le i\le n_f} ||f_i||_1^{2}\log(\min\{k,n_f\})/\varepsilon^{2} \Bigr)$, where $C_O$ is $\|O\|_{\ell_1(\mathcal{P})}^2$ or $\|O\|^2$.

\end{itemize}


To validate our theoretical framework, we perform extensive numerical simulations and a proof-of-principle experimental demonstration on a cloud-based superconducting quantum processor.
Our simulations first confirm the core protocols' validity and then showcase their utility in two distinct physical problems. We demonstrate that low-order moments, obtained with resource-efficient circuits, can yield tight bounds on the maximum eigenvalue of an unknown quantum state, a crucial quantity for error mitigation and quantum information processing.
Furthermore, by implementing our method within the QVC algorithm~\cite{Cotler2019Quantum} for a Heisenberg model, we use hybrid moments of the form $\operatorname{Tr}(O\rho^k)$ to successfully infer thermal expectation values at effectively lower temperatures.
As a key validation of its practical viability for near-term quantum computing, we execute our algorithm for measuring the second- to fourth-order Rényi entropy on a superconducting quantum device. These experiments underscore the practical advantages of our approach, providing a scalable and resource-optimal route to characterizing nonlinear properties of quantum states on near-term quantum hardware.


\section{Preliminaries}
\label{sec:tech-overview}

In this section, we establish the foundational concepts and prior art that situate our contributions. 
We begin by detailing the generalized SWAP test for estimating single moments of a quantum state. 
From there, we discuss more advanced protocols designed for polynomial state functionals, highlighting the persistent drive to reduce sample complexity. 
We then describe the current state-of-the-art, observable-weighted estimation, which achieves near-optimal sample efficiency for a broad class of problems and establishes the fundamental lower bounds for this task. 
Finally, we turn our attention from sample complexity to physical resource limitations, reviewing qubit reset as a critical technique for hardware-efficient algorithm design. 
Collectively, these topics delineate the primary frontiers in the field as the pursuit of sample-optimal methods and the practical need for resource-efficient implementations.

\subsection{Generalized SWAP Test}
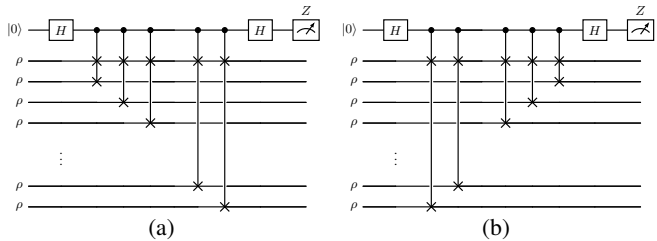
\begin{figure}[htbp]
    \centering
    \begin{tabular}{cc}
    \scalebox{0.55}{
    \begin{quantikz}
\lstick{$\ket{0}$} & \gate{H} & \ctrl{1} & \ctrl{1} & \ctrl{1} & \qw   & \ctrl{1} & \ctrl{1} & \gate{H} & \meter{Z} \\
\lstick{$\rho$}  & \qw      & \swap{1} & \swap{2}   &\swap{3} & \qw   & \swap{5} & \swap{6}   & \qw & \qw\\
\lstick{$\rho$}  & \qw      & \targX{} & \push{\,\,}  & \push{\,\,} & \qw  &\push{\,\,}& \push{\,\,} & \qw & \qw\\
\lstick{$\rho$}  & \qw      & \qw      & \targX{}   & \push{\,\,} & \qw  & \push{\,\,}& \push{\,\,} & \qw & \qw\\
\lstick{$\rho$}  & \qw      & \qw     & \qw        & \targX{} & \qw  & \push{\,\,} & \push{\,\,}& \qw & \qw\\
\wireoverride{t}  & \vdots \wireoverride{t}   &\wireoverride{t}  &\wireoverride{t}  &\wireoverride{t} &\wireoverride{t} &\wireoverride{t} &\wireoverride{t} &\wireoverride{t} \\
\lstick{$\rho$}  & \qw      & \qw      & \qw      & \qw      & \qw  &\targX{}    & \push{\,\,}& \qw & \qw\\
\lstick{$\rho$}  & \qw      & \qw      & \qw      & \qw      & \qw  & \qw    & \targX{}  & \qw & \qw
\end{quantikz}} &
\scalebox{0.55}{
\begin{quantikz}
\lstick{$\ket{0}$} & \gate{H} & \ctrl{1} & \ctrl{1} & \qw  & \ctrl{1}   & \ctrl{1} & \ctrl{1}& \gate{H} & \meter{Z} \\
\lstick{$\rho$}  & \qw      & \swap{6} & \swap{5}   &\qw & \swap{3} & \swap{2} & \swap{1}& \qw   & \qw    \\
\lstick{$\rho$}  & \qw      & \push{\,\,}& \push{\,\,} & \qw  & \push{\,\,} & \push{\,\,}& \targX{}    & \qw & \qw\\
\lstick{$\rho$}  & \qw      & \push{\,\,}      & \push{\,\,}  & \qw     & \push{\,\,}  & \targX{} & \qw       & \qw & \qw\\
\lstick{$\rho$}  & \qw      & \push{\,\,}    & \push{\,\,}       & \qw & \targX{}  & \qw   & \qw       & \qw & \qw\\
\wireoverride{t}      &\wireoverride{t} \vdots   & \wireoverride{t} & \wireoverride{t}       & \wireoverride{t}     & \wireoverride{t}& \wireoverride{t}   & \wireoverride{t}& \wireoverride{t} \\
\lstick{$\rho$}  & \qw      & \push{\,\,}      & \targX{}  & \qw      & \qw  &\qw    & \qw & \qw & \qw\\
\lstick{$\rho$}  & \qw      & \targX{} & \qw      & \qw      & \qw  & \qw    & \qw  & \qw & \qw
\end{quantikz}}\\
(a) & (b)
    \end{tabular}
    \caption{Circuits for generalized SWAP test.
    The permutation $P_k$ can be implemented as either (a) a left shift or (b) a right shift.}
    \label{fig:gen-swap-circ}
\end{figure}

The generalized SWAP test~\cite{Artur2002Direct,gitiaux2022swap} provides a direct method for estimating the $k$-th moment $\operatorname{Tr}(\rho^k)$ of an $m$-qubit quantum state $\rho$. 
Assume one can repeatedly and independently prepare $k$ registers in the identical state $\rho$, i.e., the joint state is $\rho^{\otimes k}$. 
An ancilla qubit initialized in $|0\rangle$ is appended and the controlled-permutation gate $\mathrm{C}\!-\!P_{k} = |0\rangle\langle0|\otimes\mathbb{I} + |1\rangle\langle1|\otimes P_{k}$, with $P_{k}$ the permutation of the $k$ registers, employed in the following circuit: (i) Hadamard on the ancilla; (ii) apply $\mathrm{C}\!-\!P_{k}$ (i.e., apply $P_{k}$ conditioned on the ancilla being $|1\rangle$); (iii) second Hadamard and measurement of the ancilla in the $Z$ basis.
The complete circuit is shown in Fig.~\ref{fig:gen-swap-circ}.

Mapping the measurement outcome to $b\in\{+1,-1\}$ with $+1$ ($-1$) corresponding to $|0\rangle$ ($|1\rangle$), the ancilla is observed in $|0\rangle$ with probability $p_0 = \bigl(1+\operatorname{Tr}(P_{k}\rho^{\otimes k}) \bigr)/2$. The expectation value of each outcome is $\mathbb{E}[b]=\operatorname{Tr}(P_{k}\rho^{\otimes k})$. 
Repeating the experiment independently $K$ times yields outcomes $b_1,\dots ,b_K\in\{-1,+1\}$. Define the sample mean $\hat{b}:=\frac1K\sum_{i=1}^{K} b_i$, which satisfies $\mathbb{E}[\hat{b}] = \operatorname{Tr}(P_{k}\rho^{\otimes k})$, so it is an unbiased estimator of the desired expectation value.

The utility of this estimator stems from the well-known identity $\operatorname{Tr}(P_{k}\rho^{\otimes k}) = \operatorname{Tr}(\rho^{k})$. Since the outcomes $b_i \in \{-1,+1\}$ are independent and bounded, the sample mean $\hat{b}$ concentrates around its expectation value. Specifically, Hoeffding's inequality provides a direct bound on the estimation error:
\begin{equation*}
\mathbb{P}\bigl(|\hat{b}-\operatorname{Tr}(\rho^{k})|>\varepsilon\bigr) \le 2\exp(-K\varepsilon^{2}/2).
\end{equation*}
To ensure the failure probability is bounded by a small constant $\delta$, the number of measurements must be $K \ge 2\log(2/\delta)/\varepsilon^{2}$. For a fixed $\delta$, this implies a measurement complexity of $\mathcal{O}(1/\varepsilon^{2})$. Consequently, as each measurement requires $k$ state preparations, the generalized SWAP test estimates the single moment $\operatorname{Tr}(\rho^{k})$ to additive accuracy $\varepsilon$ with a total of $\mathcal{O}(k/\varepsilon^{2})$ copies of $\rho$.

\subsection{Polynomial State Functionals}
While the generalized SWAP test is effective for estimating a single moment, many applications require the estimation of a more complex quantity: a degree-$k$ state polynomial $f(\rho) = \sum_{j=1}^{k} \alpha_j \mathrm{Tr}(\rho^j)$ with real coefficients $\alpha_j \in \mathbb{R}$. 
A central goal is to estimate $f(\rho)$ to a target additive accuracy $\varepsilon$. 
A naive strategy for this task would be to apply the generalized SWAP test to each moment $\mathrm{Tr}(\rho^j)$ individually and then compute their linear combination. 
To bound the total error by $\varepsilon$, one can estimate each term $\alpha_j \mathrm{Tr}(\rho^j)$ to an accuracy of $\varepsilon/k$, which in turn requires estimating each $\mathrm{Tr}(\rho^j)$ to an accuracy of $\varepsilon/(k|\alpha_j|)$. 
In the worst case, this is bounded by $\mathcal{O}(k^2 \|f\|_1^2 / \varepsilon^2)$.



More sophisticated paradigms exist for implementing state functionals. 
A prominent example is Quantum Signal Processing (QSP)~\cite{low2016methodology,martyn2021grand,martyn2024parallel}, which provides a versatile framework for a wide range of quantum algorithms. 
This technique is renowned for achieving provably optimal query complexity for many important quantum computing tasks. 
However, the practical application of QSP typically involves a non-trivial classical pre-processing stage, which requires finding a suitable polynomial approximation of the target function and then compiling this polynomial into a precise sequence of gate rotation angles. 
This two-stage process can introduce a notable overhead.

In contrast, the recently proposed Quantum State Function (QSF) protocol~\cite{yao2024nonlinear} offers a more direct and programmable solution specifically for polynomials. 
By integrating the Linear Combination of Unitaries (LCU) technique with the SWAP test, it coherently estimates the entire polynomial, achieving an improved sample complexity of $\mathcal{O}(k\|f\|_1^2/\varepsilon^2)$. 
A key feature of the QSF framework is that it bypasses the need for polynomial fitting and angle synthesis by establishing a direct, one-to-one mapping from the polynomial coefficients $\alpha_j$ to the circuit parameters.


Furthermore, a more general and resource-efficient goal is to design a single, coherent protocol capable of simultaneously estimating multiple distinct nonlinear functions, such as R\'{e}nyi entropy, Schatten norms, or partition functions~\cite{Wu2022Estimating, chen2024simplelowerboundcomplexity}. Such a unified method would dramatically reduce the total sample complexity required. Unfortunately, a specific and practical implementation for such a circuit does not currently exist, highlighting a key open challenge in the field.

\subsection{Observable-Weighted Estimation and its Lower Bound}
Beyond estimating intrinsic state properties, a more physically relevant paradigm emerges when incorporating an observable $O$. This generalization is essential for probing the interplay between a system's state and a specific physical quantity, such as its Hamiltonian, which is the central goal in many quantum simulation and computation tasks. This necessity gives rise to the challenge of estimating generalized moments, $\operatorname{Tr}(O\rho^j)$, and their corresponding nonlinear functionals, $f(O,\rho) = \sum_{j=1}^{k}\alpha_{j}\operatorname{Tr}(O\rho^{j})$.

In~\cite{chen2025simultaneous}, a coherent protocol is introduced to simultaneously estimate all $k$ expectations from $\mathrm{Tr}(O\rho)$ to $\mathrm{Tr}(O\rho^k)$. The approach constructs a set of $k$ Hermitian, pairwise commuting, and unbiased observables by promoting the usual cyclic-shift trick to weighted permutations and then averaging over the corresponding $S_{k}$-orbits. The favorable properties of these estimators enable a sophisticated statistical analysis that combines the union bound with median of means estimation to tightly control the overall error. This refined approach demonstrates that selecting $\Theta\bigl(k\log k \|O\|^{2}/\varepsilon^{2}\bigr)$ copies is sufficient to estimate all $\operatorname{Tr}(O\rho^{j})$ values within an additive error $\varepsilon$ with a success probability of at least $2/3$.

This result is optimal up to a logarithmic factor. To demonstrate this, one can construct a hard instance using two states that are specifically designed to be nearly indistinguishable, such as
\begin{equation*}
\rho_{\pm}= \bigl(1-\tfrac{1}{k}\pm\tfrac{\varepsilon}{k}\bigr)\lvert0\rangle\langle0|
+\bigl(\tfrac{1}{k}\mp\tfrac{\varepsilon}{k}\bigr)\lvert1\rangle\langle1|.
\end{equation*}
For these states, distinguishing the value of a single high-order moment like $\operatorname{Tr}\bigl(O\rho^{k}\bigr)$ becomes challenging. By applying the Helstrom--Holevo bound~\cite{HELSTROM1967Detection, HOLEVO1973Statistical} to this scenario, one can show that any unbiased estimation procedure requires at least $\Omega\bigl(k\|O\|^{2}/\varepsilon^{2}\bigr)$ copies~\cite{chen2025simultaneous}. Hence, the simultaneous estimation protocol is shown to be nearly optimal, incurring only an additional $\log k$ factor in sample cost compared to the fundamental limit imposed by the most challenging individual moments. However, while~\cite{chen2025simultaneous} offers a complete theoretical proof of sample efficiency, it does not provide a concrete method for its implementation. Therefore, how to design an efficient quantum circuit for this powerful protocol remains a crucial open question, which is the central focus of our work.


\subsection{Hardware Efficiency and Qubit Reset}

Beyond sample complexity, the number of physical qubits required is a critical bottleneck for implementing advanced quantum algorithms on near-term hardware. As established in the preceding sections, protocols like the generalized SWAP test require access to $k$ distinct state registers, leading to a circuit width that scales with the number of moments to be estimated (e.g., $km+1$ qubits). This scaling can be prohibitively large for many applications of interest.

A key improvement was introduced in~\cite{yirka2021qubit}, leverages qubit reset techniques to reduce the circuit width from $km+1$ to $2m+1$. The viability of such resource-efficient protocols, however, hinges entirely on the ability to perform fast and high-fidelity qubit resets in practice.

The physical implementation of such a reset operation faces significant challenges. According to Landauer's principle, there is a fundamental thermodynamic minimum for the energetic cost of erasing one bit of information~\cite{Landauer1961irreversibility,bennett1982thermodynamics, georgescu202160}. In any realistic device, the energy consumed often exceeds this theoretical limit by several orders of magnitude. This discrepancy arises because the quasi-static regime assumed by Landauer's principle is incompatible with quantum operations that must satisfy stringent demands for both high speed and high fidelity~\cite{liu2025optimally}. This inherent conflict between energy, speed, and fidelity poses a significant barrier to the scalability and efficiency of quantum computing.

To address this implementation bottleneck, finite-time qubit reset has become a major focus of recent research. These efforts span from deriving universal thermodynamic inequalities that bound the cost of erasure~\cite{Zhen2021Universal, Van2022finite, zhen2022inverse} to identifying optimal, energy-efficient protocols for specific physical systems~\cite{Magnard2018fast, zhou2021rapid}. Driven by the practical need for reset operations of extremely high speed and fidelity~\cite{Wang2024Efficient}, a deep understanding and mastery of this fundamental operation is of paramount strategic importance. 

\section {Main Results}

\subsection{Resource-efficient circuit for simultaneous moment estimation}
\label{sec:moment_result}
Here, we present the first explicit circuit capable of simultaneously estimating all moments up to order $k$, featuring $k$-independent width and $\mathcal{O}(k)$ depth, with sample complexity scaled as $\mathcal{O}(k\log k)$. For an $m$-qubit state $\rho$, we reduce the total number of qubits to $2m+1$ and the total number of SWAP gates to $\mathcal{O}(k)$ by leveraging qubit reset techniques. We summarize the main theoretical guarantee as follows:


\begin{theorem}
(Sample-efficient simultaneous moment estimation) 
Let $\rho$ be an arbitrary $m$-qubit quantum state. 
For any positive integer $k$, all moments up to order $k$, $\operatorname{Tr}(\rho^{2}), \dots, \operatorname{Tr}(\rho^{k})$, can be estimated simultaneously using a circuit on $2m + 1$ qubits with depth $\mathcal{O}(k)$. 
This estimation achieves an additive error $\varepsilon$ with probability at least $2/3$, using $\mathcal{O}\bigl(k \log k/ \varepsilon^{2}\bigr)$ copies of $\rho$.
\label{thm:theorem_1}
\end{theorem}

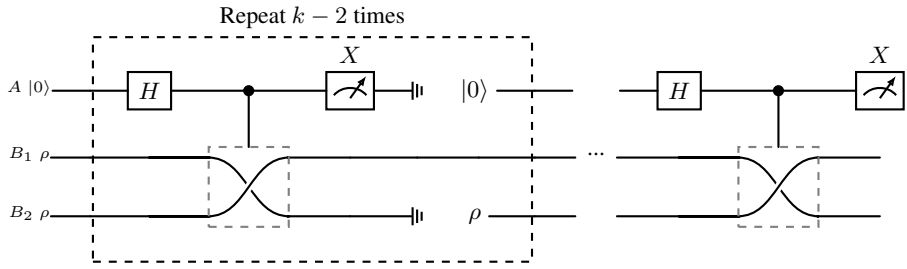
\begin{figure*}[htbp]
\centering
\scalebox{1.0}{
\begin{quantikz}[transparent]
\qw^{A \; \ket{0}}&&\gate{H}\gategroup[3,steps=5,style={inner sep=10pt,dashed,label={above:{Repeat $k-2$ times}}}]{}& \ctrl{1}& \meter{X} & \ground{} & \wireoverride{n} \push{\ket{0}\,\,}& & \midstick[3,brackets=none]{...}&
\gate{H}&\ctrl{1}& \meter{X} \\
\qw^{B_1 \; \rho} &&  &\gate[2,style={draw=gray, dashed}]{\permute{2,1}}&&&&&&
&\gate[2,style={draw=gray, dashed}]{\permute{2,1}}&\\
\qw^{B_2 \; \rho} & & && &\ground{} & \wireoverride{n} \push{\rho\,\,}&&&
&&
\end{quantikz}
}\\
\caption{Quantum circuit incorporating qubit reset operations. 
Here, after measurement, the auxiliary qubits are reset to \(\ket{0}\), while the register $B_2$ storing \(\rho\) is directly reset and re-prepared in the state \(\rho\). 
All the measurement outcomes are recorded for post-processing to obtain all the moments.
This strategy reduces the overall qubit requirement to \(2m+1\), significantly lowering the hardware overhead while preserving the estimation capability.
}
\label{fig:theorem1_moment}
\end{figure*}

We defer the detailed proof of Theorem~\ref{thm:theorem_1} to Appendix~\ref{appendix:proof_of_theorem_1}, and the corresponding circuit is depicted in Fig.~\ref{fig:theorem1_moment}. Unlike previous approaches that employ a different circuit for each moment $\operatorname{Tr}(\rho^{j})$, our protocol uses a single fixed circuit and extracts all moments through a joint measurement and classical post-processing, significantly streamlining the experimental workflow. The circuit depth and the number of CSWAP gates scale linearly with $k$ ($\mathcal{O}(k)$), matching the standard SWAP-test required for estimating only the $k$-th moment.

The qubit overhead is minimized by a dual-reset strategy. First, following the qubit-recycling principle introduced in~\cite{yirka2021qubit}, the data register that held~$\rho$ is actively re-initialized to $\lvert0\rangle$ after every execution. This allows the same hardware qubits to host a freshly prepared copy of $\rho$ in the next round. Concurrently, the $k-1$ auxiliary qubits are consecutively measured in the $X$-basis followed by resetting to their ground state~\cite{decross2023qubit}, making them immediately reusable. While this combined reset mechanism does not alter the fundamental sample complexity $\mathcal{O}(k\log k/\varepsilon^{2})$, it reduces the physical qubit count to $2m+1$, independent of~$k$.

Algorithm~\ref{alg:theorem1_moment} summarizes the measurement routine. A quantitative comparison with prior schemes is provided in Table~\ref{tab:resource-comparison}, covering qubit count, circuit depth, CSWAP gate number, and sample complexity. 
Notably, whereas earlier protocols need $\mathcal{O}(k^{2}/\varepsilon^{2})$ copies to estimate all $k$ moments, our method lowers the sample complexity to $\mathcal{O}(k\log k/\varepsilon^{2})$. This represents a quadratic improvement and is only a logarithmic factor away from the best existing lower bound, while remaining competitive in every other resource metric.



\begin{algorithm}[htbp]
\SetAlgoLined
  \KwData{the circuit in Fig.~\ref{fig:theorem1_moment}, $n\geq \mathcal{O}(k\log k/\varepsilon^2)$.}
  \KwResult{estimate for $\mathrm{Tr}(\rho^2)$, $\dotsc$, $\mathrm{Tr}(\rho^k)$.}
    $p_1 \leftarrow 0$, $\dotsc$, $p_{k-1} \leftarrow 0$.\\
    \For{$i=1$ to $n$}{
    Execute the circuit and record the outcomes of the measurements as $x_1$, $\dotsc$, $x_{k-1}$ sequentially, where $x_j\in \{1, -1\}$ for $j=1,\dotsc, k-1$.\\
        \For{$l=1$ to $k-1$}{
        $p_l\leftarrow \prod_{j=1}^{l}x_j$.
        }
    }
    $p_1\leftarrow p_1/n$, $\dotsc$, $p_{k-1}\leftarrow p_{k-1}/n$.\\
    Return $p_j$ as estimate for $\mathrm{Tr}(\rho^{j+1})$ for $j=1, \dotsc, k-1$.
    \caption{Algorithm for moment estimation.}
    \label{alg:theorem1_moment}
\end{algorithm}

\begin{table*}[htbp]
  \centering
  \caption{Resource Comparison for Quantum State Moment Estimation Methods. 
  Here we compare our method with existing approaches in terms of qubit count, circuit depth, CSWAP gate count, and sample complexity for estimating all the quantum state moments $\{\operatorname{Tr}(\rho^i)\}_{i=1}^k$. 
  Here, $r$ denotes the rank of $\rho$. 
  The ``Original $|\psi\rangle$'' column specifies whether an algorithm needs access to the purifying pure state $|\psi\rangle$ that generates $\rho$, or if it can work directly with copies of the mixed state $\rho$.}
  \label{tab:resource-comparison}
  \begin{tabular}{lccccc}
    \toprule
    Method                 &  Qubits             &  Depth  & CSWAP Gates    &  Copies of $\rho$ & Original $|\psi\rangle$\\
    \midrule
    SWAP Test~\cite{Artur2002Direct}              & \(\mathcal{O}(k)\)  & \(\mathcal{O}(k)\)  & \(\mathcal{O}(k)\)  & \(\mathcal{O}\bigl(k^2/\varepsilon^2\bigr)\) & Not required\\[2mm]
    Two-copy Test~\cite{subacsi2019entanglement}          & \(\mathcal{O}(k)\)  & \(\mathcal{O}(k)\)  & \(\mathcal{O}(k)\)  & \(\mathcal{O}\bigl(k^2/\varepsilon^2\bigr)\) & Required\\[2mm]
    SWAP test + Reset~\cite{yirka2021qubit}      & \(\mathcal{O}(1)\)  & \(\mathcal{O}(k)\)  & \(\mathcal{O}(k)\)  & \(\mathcal{O}\bigl(k^2/\varepsilon^2\bigr)\) & Not required\\[2mm]
    TCT test + Reset~\cite{yirka2021qubit}       & \(\mathcal{O}(1)\)  & \(\mathcal{O}(k)\)  & \(\mathcal{O}(k)\)  & \(\mathcal{O}\bigl(k^2/\varepsilon^2\bigr)\) &  Required\\[2mm]
    Multivariate Trace~\cite{quek2024multivariate}     & \(\mathcal{O}(1)\)  & \(\mathcal{O}(k)\)  & \(\mathcal{O}(k)\)  & \(\mathcal{O}\bigl(k^2/\varepsilon^2\bigr)\) & Not required\\[2mm]
    Rank Estimating~\cite{shin2024rank} & \(\mathcal{O}(r)\) & \(\mathcal{O}(1)\) &\(\mathcal{O}(r)\) & \(\mathcal{O}\bigl(k^2/\varepsilon^2\bigr)\) & Not required\\[2mm]
    \midrule
    \textbf{This Work}           & \textbf{\(\mathcal{O}(1)\)}  & \(\mathcal{O}(k)\)  & \(\mathcal{O}(k)\)  & \textbf{\(\mathcal{O}\bigl(k\log k/\varepsilon^2\bigr)\)} & Not required\\[1mm]
    \bottomrule
  \end{tabular}
\end{table*}

\subsection{Nonlinear functional of quantum state}

In Theorem~\ref{thm:theorem_1} we first estimate every moment $\operatorname{Tr}(\rho^{j})$ and only then form a polynomial $f(\rho)=\sum_{j=1}^{k}\alpha_{j}\operatorname{Tr}(\rho^{j})$.
For many practical tasks, one is interested solely in the numerical value of $f(\rho)$ rather than the full moment vector. 
This can be achieved easily through Theorem~\ref{thm:theorem_1} and classical post-processing.
Here we present an efficient estimate of $f(\rho)$ that does not require any classical post-processing.
Suppose we have $p_1$, $\dotsc$, $p_{k-1}$ from Algorithm~\ref{alg:theorem1_moment} with $n=\mathcal{O}(k\|f\|_1^2/\varepsilon^2)$.
Denote $\hat{p}=\sum_{j=2}^{k}\alpha_{j}p_{j-1} + \alpha_{1}$.
We have $\mathbb{E}[\hat{p}] = f(\rho)$ and 
\[
\mathrm{Var}[\hat{p}]=\mathrm{Var}\biggl[\sum_{j=2}^{k}\alpha_{j}p_{j-1}\biggr]\leq \biggl(\sum_{j=2}^{k}|\alpha_j|\biggr)^2\leq \|f\|_1^2,
\]
since $\mathrm{Var}[p_j]\leq 1$ follows from Appendix~\ref{appendix:proof_of_theorem_1}.
By Chebyshev's inequality, we can obtain an estimate for $f(\rho)$ within additive error $\varepsilon$ with probability at least $2/3$.

However, while provably efficient, this approach has practical drawbacks. It necessitates the measurement of a set of $k$ commuting observables, followed by a classical post-processing step to combine the results. A more streamlined and elegant approach would be to engineer a single quantum circuit whose output directly corresponds to the value of $f(\rho)$, obviating the need for intermediate moment estimation and classical processing entirely. In what follows, we present such a method. We design a quantum circuit that evaluates $f(\rho)$ by mapping its value onto the expectation value of a single Pauli operator on one ancillary qubit, $\langle X \rangle = f(\rho)$. This offers a significant simplification in the measurement and data processing workflow. Specifically, we state our theorem as follows.

\begin{theorem}(Resource-efficient evaluation of a polynomial state functional)
\label{thm:theorem_2}
Let $\rho\in\mathbb{C}^{2^{m}\times 2^{m}}$ be an arbitrary $m$-qubit state and let $f(\rho)$ be the degree-$k$ polynomial state functional defined previously.
For any $\varepsilon>0$, $f(\rho)$ can be estimated using a circuit on $2m+\lceil\log_{2}\! k\rceil+1$ qubits within additive error $\varepsilon$ with success probability at least $2/3$, using $\mathcal{O}\bigl(k\|f\|_{1}^{2}/\varepsilon^{2}\bigr)$ copies of $\rho$.
The circuit employs $\mathcal{O}(k)$ CSWAP gates, $\mathcal{O}(k)$ qubit-reset operations, and requires only single-qubit measurements.
\end{theorem}

The proof of Theorem~\ref{thm:theorem_2} is detailed in Appendix~\ref{appendix:proof_of_theorem_2}. A complete implementation is detailed in Algorithm~\ref{alg:theorem2_singleqsf}, with the gate-level diagram provided in Appendix~\ref{appendix:proof_of_theorem_2}. 
In contrast to other methods, QSF offers significant advantages in both experimental feasibility and programmability. 
It operates directly on standard, identical copies of the quantum state, thus obviating the need for the experimentally challenging purified query access required by QSP-based frameworks. 
Furthermore, QSF provides a direct and explicit embedding of polynomial coefficients into circuit parameters, which circumvents the complex function-fitting and angle-synthesis steps inherent to alternative approaches, thereby enabling the straightforward implementation of arbitrary polynomials.

Building upon these foundational advantages, our work introduces a series of critical optimizations to the QSF protocol that substantially reduce its resource overhead and enhance its practicality for near-term quantum hardware. 
Our primary optimization achieves a quadratic reduction in gate complexity. 
By engineering the circuit as a recursive SWAP-test structure, the cost of measuring each subsequent moment, $\operatorname{Tr}(\rho^{j+1})$, is reduced to a single additional CSWAP gate. 
This innovation reduces the total CSWAP gate count from a prohibitive $\mathcal{O}(k^2)$ to a far more scalable $\mathcal{O}(k)$.

Complementing this, we introduce a significantly simplified coefficient-handling mechanism. Whereas the original protocol requires $k$ controlled-$R_y$ gates to manage signs, our design handles the common case of non-negative coefficients with just a single Hadamard gate, while instances with negative coefficients require only a single, targeted CZ gate. These optimizations yield a circuit that is not only shallower but also inherently more robust to noise, a crucial advantage for achieving high fidelity on contemporary quantum processors. A detailed comparison between our method and the original QSF protocol is presented in Table~\ref{tab:qsf-comparison}.

\begin{table*}[htbp]
  \centering
  \caption{Resource comparison for estimating a polynomial state functional $f(\rho)$. 
  We compare our optimized protocol with the original QSF method~\cite{yao2024nonlinear}. 
  The ``Control State Preparation'' column refers to the subroutine used to generate the control states for selecting the powers of $\rho$. 
  The ``Sign Gates'' column details the gates required to encode the signs of the coefficients $\alpha_j$.}
  \label{tab:qsf-comparison}
  \begin{tabular}{lccccc}
    \toprule
    Method   &  Qubits   &  Depth  & CSWAP Gates & Control State Preparation & Sign Gates \\
    \midrule
    QSF~\cite{yao2024nonlinear} & $km +\lceil\log_2\! k \rceil + 1$ & \(\mathcal{O}(k^2)\) & \(\mathcal{O}(k^2)\) & Arbitrary state preparation & $k$ controlled-$R_y$  \\
    Ours           & $2m +\lceil\log_2\! k \rceil + 1$ & \(\mathcal{O}(k)\)  & \(\mathcal{O}(k)\) & Gray code sequence & $\le k/2$ CZ \\
    \bottomrule
  \end{tabular}
\end{table*}


\begin{algorithm}[htbp]
\SetAlgoLined
  \KwData{The circuit in Fig.~\ref{fig:QSF_new_circuit} (b) and $n\geq \mathcal{O}\bigl(\|f\|_1^{2}/\varepsilon^{2}\bigr)$.}
  \KwResult{estimate for $f(\rho)=\sum_{j=1}^{k}\alpha_{j}\operatorname{Tr}\bigl(\rho^{j}\bigr)$.}
    $P \leftarrow 0$, $\lambda_i\leftarrow |\alpha_{i}|/\|f\|_1$, for $i=1, \dotsc, k$.\\
    Generate the gray code for $\{0, 1,\dotsc, k-1 \}$, denote the Gray code for $i$ as $g(i)$.\\
    \For{$i=1$ to $k-1$}{
    $\Lambda_i \leftarrow \sum_{j=i}^k \lambda_j$.\\
    $\theta \leftarrow 2\arccos \sqrt{\lambda_i / \Lambda_i}$.\\
    Let $t_i$ denote the bit that is different between $g(i-1)$ and $g(i)$. 
    Initialize gate $V_i$ as controlled-$R_y(\theta)$, where the target is $B_{t_i}$ and the control system is system $B$ except $B_{t_i}$.
    The condition is $g(i)$ except for the $t_i$ bit.\\
    }
    \For{$i=1$ to $k$}{
        \If{$\alpha_i <0$ and $\sum_{j=1}^{k}\mathrm{sgn}(\alpha_j) \geq 0$ or $\alpha_i >0$ and $\sum_{j=1}^{k}\mathrm{sgn}(\alpha_j) < 0$}{
        Add $\ket{i-1}$ to the negative basis.\\
        }
    }
    \For{$i=1$ to $n$}{
    Execute the circuit and record the outcomes of the measurements as $x\in\{1, -1 \}$.\\
    $P \leftarrow P +x$.\\
    }
    $P\leftarrow \|f\|_1P/n$.\\
    \If{$\sum_{j=1}^{k}\mathrm{sgn}(\alpha_j) < 0$}{
    $P\leftarrow -P$.\\
    }
    Return $P$ as estimate for $f(\rho)$.
    \caption{Algorithm for quantum state function.}
    \label{alg:theorem2_singleqsf}
\end{algorithm}

\begin{figure*}[htbp]
    \centering
        \includegraphics[width=0.75\linewidth]{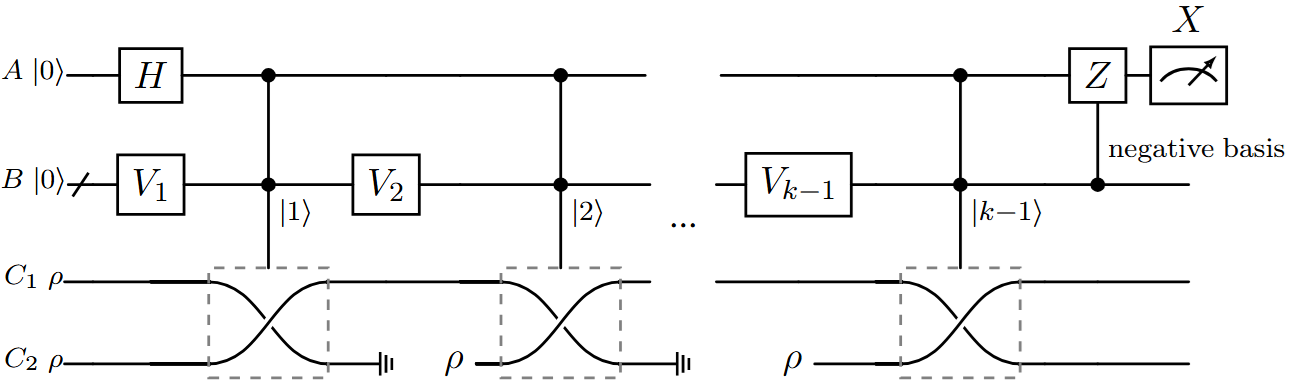}
    \caption{Depth-optimized quantum circuit for estimating the trace-polynomial \(f(\rho)\) of degree $k$.
    The circuit implements a technique known as CSWAP-based data re-uploading. The core of this method involves sequentially encoding the polynomial's coefficients onto an auxiliary qubit using a set of controlled operators ${V_j}$, which are implemented as Givens rotations, each followed by a controlled-SWAP gate.} For a detailed construction of the $V_j$ operators, please see Appendix~\ref{appendix:proof_of_theorem_2}. The small mark close to the control system denotes the control condition.
    \label{fig:QSF_new_circuit}
\end{figure*}


In practical applications, a single quantum state functional is insufficient to capture all properties of interest about the quantum state. 
To obtain a faithful and operationally complete description one therefore wants to evaluate an ensemble of distinct polynomial functionals $f_{1}(\rho)$, $f_{2}(\rho)$, $\dotsc$, $f_{n_{f}}(\rho)$, each of which may involve a different collection of moments up to some order \(k_{i}\).  
Naively applying the single functional protocol of Theorem~\ref{thm:theorem_2} sequentially to each \(f_i\) would require \(n_f\) times as many copies of \(\rho\) and similarly increase the gate overhead. 
However, since all these functions operate on the same input state, the state \(\rho\) can be reused across the different evaluations. 
This reuse effectively reduces the overall sample complexity needed to compute the desired estimates. Therefore, by combining Theorem~\ref{thm:theorem_1}, Theorem~\ref{thm:theorem_2}, and the above analysis, we derive the following corollary.


\begin{corollary}(Resource-efficient evaluation of multi-polynomial functionals)
\label{cor:multi-funcs}
Let $\rho$ be an $m$-qubit state. 
Consider a set of $n_f$ polynomial state functionals, $\{f_i(\rho)\}_{i=1}^{n_f}$. 
Each functional $f_i$ has its own degree $k_i$ and coefficients $\alpha_{i,j}$. Let $k:= \max_{1\leq i \leq n_f}k_i$ be the maximum degree in the set.
For any $\varepsilon>0$, the values $f_1(\rho),\dots,f_{n_f}(\rho)$ can be estimated within an additive error $\varepsilon$ with probability at least $2/3$ using $\mathcal{O}\Bigl(k\max_{1\le i\le n_f} ||f_i||_1^{2}\log(\min\{k,n_f\})/\varepsilon^{2} \Bigr)$ copies of $\rho$. 
\end{corollary}

The detailed circuit implementation and formal proof for this corollary are provided in Appendix~\ref{appendix:proof_of_theorem_3}. 
The core of our approach is a resource-efficient parallelization strategy that allows for the simultaneous estimation of all $n_f$ functionals. Instead of running separate experiments, we construct a single, larger quantum circuit that leverages shared access to the input state $\rho$ while dedicating separate ancilla registers to each functional $f_i$.

More specifically, our construction extends the single-functional circuit shown in Fig.~\ref{fig:QSF_new_circuit}. 
While the $2m$ qubits holding the purification of the input state $\rho$ are shared across the entire computation, we introduce $n_f$ parallel sets of ancilla qubits, each responsible for estimating a single functional $f_i$. 
The total qubit count for this unified circuit is $2m + n_f\lceil\log_2\! k\rceil + n_f$. 
The sample complexity, which scales as $\mathcal{O}\bigl(k\max_j\|f_j\|_1^{2}\log(\min\{k,n_f\})/\varepsilon^{2}\bigr)$, features a favorable logarithmic dependence on the number of functionals, $\log n_f$. This term arises from the statistical overhead of ensuring that all $n_f$ estimates are simultaneously within the desired error $\varepsilon$ with high probability. This unified, parallel strategy is what yields the highly resource-efficient evaluation of multi-polynomial state functionals presented in the corollary.

\subsection{Generalization of Nonlinear Quantum State Functions to Observables}

In the foregoing discussion, we have presented a construction of nonlinear functionals of a quantum state $\rho$.  
In many practical contexts, however, the quantities of interest are not the global moments $\operatorname{Tr}(\rho^{j})$ themselves, but rather the observable-weighted moments $\operatorname{Tr}(\hat O\rho^{j}),$ where $\hat O$ is an $m$-qubit Hermitian operator.  
We analyze below two concrete schemes for estimating the multi-polynomial family
\begin{equation*}
\begin{aligned}
    f_i(\hat O, \rho)&=\sum_{j=1}^{k_i}\beta_{i,j}    \operatorname{Tr}(\hat O\rho^{j}), 
\end{aligned}
\end{equation*}

where $k=\max_i k_i$ is the maximum degree among all polynomials and $\lVert f_i\rVert_1=\sum_{j}|\beta_{i,j}|$ denotes the L1-norm of the coefficients. Next we present two distinct strategies with a clear trade-off between sample complexity and implementation overhead. 
The first scheme is more copy-efficient but requires a more complex circuit capable of implementing controlled-oracle operations. 
The second scheme, while consuming more copies of $\rho$, is considerably easier to realize on near-term hardware as it only requires decomposing $\hat{O}$ and performing standard Pauli measurements.

The first scheme leverages the LCU technique. 
Its foundation is the Russo--Dye theorem~\cite{russo1966note}, which ensures that any operator with a spectral norm at most one can be expressed as a convex combination of unitary operators. 
For a generic observable $\hat O$ define the normalized operator $\hat O'=\hat O/\lVert\hat O\rVert$ and set  
\begin{equation*}
    U \;=\; \hat O' + \mathrm{i}\sqrt{I-\hat O'^{2}},
\end{equation*}
we have $\hat O =\lVert\hat O\rVert\bigl(U+U^{\dagger}\bigr)/2$.
This linear combination can be implemented using a single ancillary qubit via the standard LCU protocol, augmenting either the circuit from Fig.~\ref{fig:theorem1_moment} or Fig.~\ref{fig:QSF_new_circuit}. 
The ancilla is prepared in the state $(\lvert0\rangle+\lvert1\rangle)/\sqrt2$ and used to conditionally apply either $U$ or $U^{\dagger}$ to the system. 
A final measurement of the $X$ operator on this ancilla yields an expectation value proportional to $\operatorname{Tr}(\hat{O}'\rho^{j})$.

\begin{figure*}[htbp]
    \centering
        \includegraphics[width=0.75\linewidth]{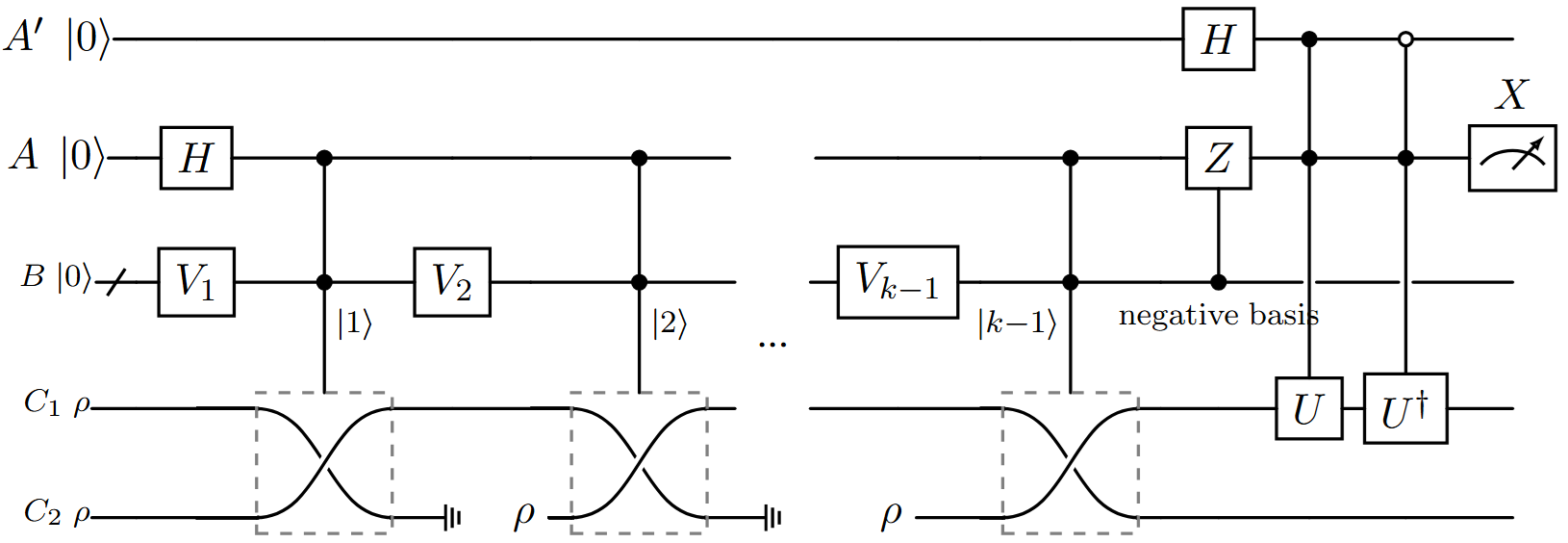}
    \caption{The quantum circuit diagram for estimating the quantity $f(O ,\rho^k)$. This architecture generalizes the circuit for estimating the trace-polynomial $f(\rho)$ by introducing an auxiliary ancilla qubit (topmost wire).}
    \label{fig:LCU_one_ancilla}
\end{figure*}

A crucial consequence of this approach is the cost associated with post-processing. 
To recover the unnormalized value $\operatorname{Tr}(\hat O\rho^{j})$, the estimator must be rescaled by the factor $\lVert\hat O\rVert$. 
This classical rescaling amplifies the statistical variance by $\lVert\hat O\rVert^{2}$, thereby increasing the required number of measurements. The total sample complexity thus becomes:
\begin{equation}\label{eq:LCU_complexity}
    N_{\mathrm{LCU}}
 \;=\;
 \mathcal{O}\Bigl(
     k\lVert\hat O\rVert^{2}
     \max_{i}\lVert f_i\rVert_{1}^{2}\log\bigl(\min\{k,n_f\}\bigr)/\varepsilon^{2}
 \Bigr).
\end{equation}
A representative circuit-level implementation is shown in Fig.~\ref{fig:LCU_one_ancilla}.
The total qubit count increases by one due to the LCU ancilla.

A more hardware-friendly alternative expands the observable in the Pauli basis
\begin{equation*}
    \hat O=\sum_{p=1}^{n_P}\alpha_p P_p, \qquad P_p\in\{I,X,Y,Z\}^{\otimes m},
\end{equation*}
and performs single-shot Pauli measurements on each copy of $\rho$.
Importance sampling proceeds as follows:
(i) Select $P_p$ with probability $p_p=|\alpha_p|/S$, where $S=\|O\|_{\ell_1(\mathcal{P})}$.
(ii) Measure $P_p$ simultaneously on every subsystem that participates in $\rho^{j}$, obtaining $m_p\in\{\pm1\}$.
(iii) Output the random variable $X=S\operatorname{sgn}(\alpha_p)m_p$.
One readily verifies $\mathbb{E}[X]=\operatorname{Tr}(\hat O\rho^{j})$, $\operatorname{Var}[X]\le S^{2}$, whence
\begin{equation}\label{eq:Pauli_complexity}
    N_{\mathrm{Pauli}}  =\mathcal{O}\Bigl(  k\|O\|_{\ell_1(\mathcal{P})}^{2} \max_i\lVert f_i\rVert_{1}^{2}\       \log\bigl(\min\{k,n_f\}\bigr)/\varepsilon^{2}     \Bigr). 
\end{equation}

Because Pauli strings form an orthonormal basis, then applying the Cauchy-Schwarz inequality and standard matrix norm inequalities to the Pauli decomposition yields:
\begin{equation*}
    \lVert\hat O\rVert\;\le\;S\;\le\;\sqrt{n_P}\,\lVert\hat O\rVert. 
\end{equation*}

For sparse observables ($n_P \ll 4^{m}$) one typically finds $S=\Theta(\lVert\hat O\rVert)$, so Eq.~\ref{eq:LCU_complexity} and Eq.~\ref{eq:Pauli_complexity} coincide up to constant factors; for highly non-local operators $S$ may scale as $n_P\lVert\hat O\rVert$.  
In many ubiquitous Hamiltonians, such as the transverse-field Ising or the Heisenberg model, $S$ and $\lVert\hat O\rVert$ differ only by an $\mathcal{O}(1)$ constant. 
Therefore, we can derive the following corollary:
\begin{corollary}(Polynomial state observables)
\label{cor:poly-observables}
Consider a set of $n_f$ polynomial observables of the form $f_i( O,\rho)=\sum_{j=1}^{k_i}\beta_{i,j}\mathrm{Tr}(O\rho^{j})$ for a Hermitian observable $\hat O$ and a state $\rho$. 
Let $k := \max_i k_i$, and let $C_O$ be the resource cost for implementing $O$, given by $C_O=\lVert O\rVert$ for an LCU decomposition or $C_O=\|O\|_{\ell_1(\mathcal{P})}$ for a Pauli decomposition.
For any $\varepsilon>0$, every value $f_i(O,\rho)$ can be estimated to additive error $\varepsilon$ with success probability at least $2/3$ using
\[
N_{\mathrm{copies}}
= \mathcal{O}\Bigl(
      kC_O^{2}
      \max_i \lVert f_i\rVert_{1}^{2} \log(\min\{k,n_f\})/\varepsilon^{2}
  \Bigr)
\]
copies of the state $\rho$.

\end{corollary}

For estimating nonlinear properties of an $m$-qubit state, such as R\'{e}nyi entropies, the primary alternatives are classical shadow and randomized measurement protocols. 
These methods, however, suffer from resource costs that scale exponentially with the number of qubits. 
For instance, estimating a generic nonlinear functional requires a sample complexity that scales with the Hilbert space dimension $d=2^m$.
That is at least $\Omega(d/\varepsilon^2)$ for classical shadows if the observable is global~\cite{huang2020predicting}. 
Furthermore, both protocols typically necessitate deep circuits to implement global random operations.

Our protocol, by contrast, fundamentally circumvents this exponential scaling. 
Its sample complexity is independent of the Hilbert space dimension, depending only polynomially on the order $k$ of the desired functional. 
In the highly relevant near-term regime where $k \ll d$, our method's complexity of $\mathcal{O}\bigl(\text{poly}(k)/\varepsilon^2\bigr)$ provides an exponential improvement in the number of qubits, $m$, over existing approaches. 
This advantage, realized by the circuit construction in Corollary~\ref{cor:poly-observables}, makes our protocol exceptionally well-suited for practical applications on near-term quantum computers.

\section{Experimental Evaluation}

All experiments in this section were conducted on two different servers. 
The first server is equipped with an AMD EPYC 9654 96-Core CPU, two NVIDIA GeForce RTX 4090 GPUs, and 503~GB of RAM. The second server is equipped with an Intel Core i9-14900K CPU, a single NVIDIA GeForce RTX 4090 GPU, and 125~GB of RAM.

\subsection{Estimating the maximum eigenvalue from low-order moments}
\label{subsec:max_eigenvalue_estimation}

Let $0\le\lambda_1\le\lambda_2\le\dots\le\lambda_{2^{m}}=\lambda_{\max}$ be the spectrum of $\rho$. 
The eigenvalue spectrum of a quantum state governs a wide range of information-theoretic tasks~\cite{somma2019quantum, Nghiem2023quantum}. 
In particular, the maximum eigenvalue $\lambda_{\max}$ encodes the min-entropy $H_{\min}(\rho)=-\log\lambda_{\max}$~\cite{Konig2009Entropy}, determines compression rates, and enters protocols such as quantum PCA~\cite{lloyd2014quantum, Tao2021experimental}, error-mitigation~\cite{Cai2023quantum}, and distillation~\cite{Campbell2012magicstate, Bravyi2012magicstate}. 
In this work, we present a framework to obtain an interval for $\lambda_{\max}$.
Unlike previous approaches based on quantum PCA or exact moment estimation~\cite{Tanaka2014determining}, our method utilizes low-order moments subject to sampling noise, as higher-order moments exhibit increased sensitivity to noise.
This strategy aligns with the constraints of noisy intermediate-scale quantum (NISQ)~\cite{preskill2018quantum,chen2023complexity} or early fault-tolerant quantum computing (EFTQC)~\cite{Katabarwa2024early, Liang2024modeling}.

Suppose we have unbiased estimates $\hat m_j$ of the moments $m_j=\mathrm{Tr}(\rho^{j})=\sum_{i}\lambda_i^{j}$ for $j=2,\dots,k$, each with additive error at most $\varepsilon$, i.e. $|m_j-\hat m_j|\le\varepsilon$. 
We derive the following interval for $\lambda_{\max}$:
\begin{equation}
\begin{aligned}
\biggl[&\max\biggl\{\max_{2\le j\le k-1} \frac{\hat m_{j+1}-\varepsilon}{\hat m_j+\varepsilon}, \max_{2\le j\le k} (\hat m_{j}+\varepsilon)^{1/(j-1)} \biggr\} , \\
&\min_{2\le j\le k}(\hat m_j+\varepsilon)^{1/j}\biggr]
\end{aligned}
\label{eq:lambda-max-interval}
\end{equation}

The proof of Eq.~\eqref{eq:lambda-max-interval} can be found in Appendix~\ref{appendix:proof_of_interval}.
We perform the following Monte Carlo experiment ($1000$ trials per setting):
\begin{enumerate}
  \item Fix a rank \(n\in\{2, 4, 8, \dotsc, 32\}\) and choose   $\lambda_1,\dots,\lambda_n\sim\text{Dirichlet}(1,\dots,1)$.
  \item For $j=2,3,4$ set
        $\hat m_j = m_j + \mathrm{sgn}(\xi_j)\min\{ \varepsilon, |\xi_j|\}$ with
        $\xi_j\sim\mathcal{N}(0,\varepsilon^{2}/4)$.
  \item Compute the interval and record its absolute length. 
\end{enumerate}

The length of the resulting interval directly characterizes the tightness of the bound.  
For instance, with an interval length of $0.1$, the largest eigenvalue $\lambda_{\max}$ can be constrained within an additive error of $0.05$, which is negligible compared to the full range $[0,1]$.  
In addition, the derivation relies only on elementary inequalities that can be evaluated efficiently on a classical computer.  
With more sophisticated algorithmic or numerical techniques, the interval length could potentially be reduced even further.  
A systematic exploration of such improvements constitutes an interesting direction for future research.

\begin{figure}[t]
  \centering
       \includegraphics[width=\linewidth]{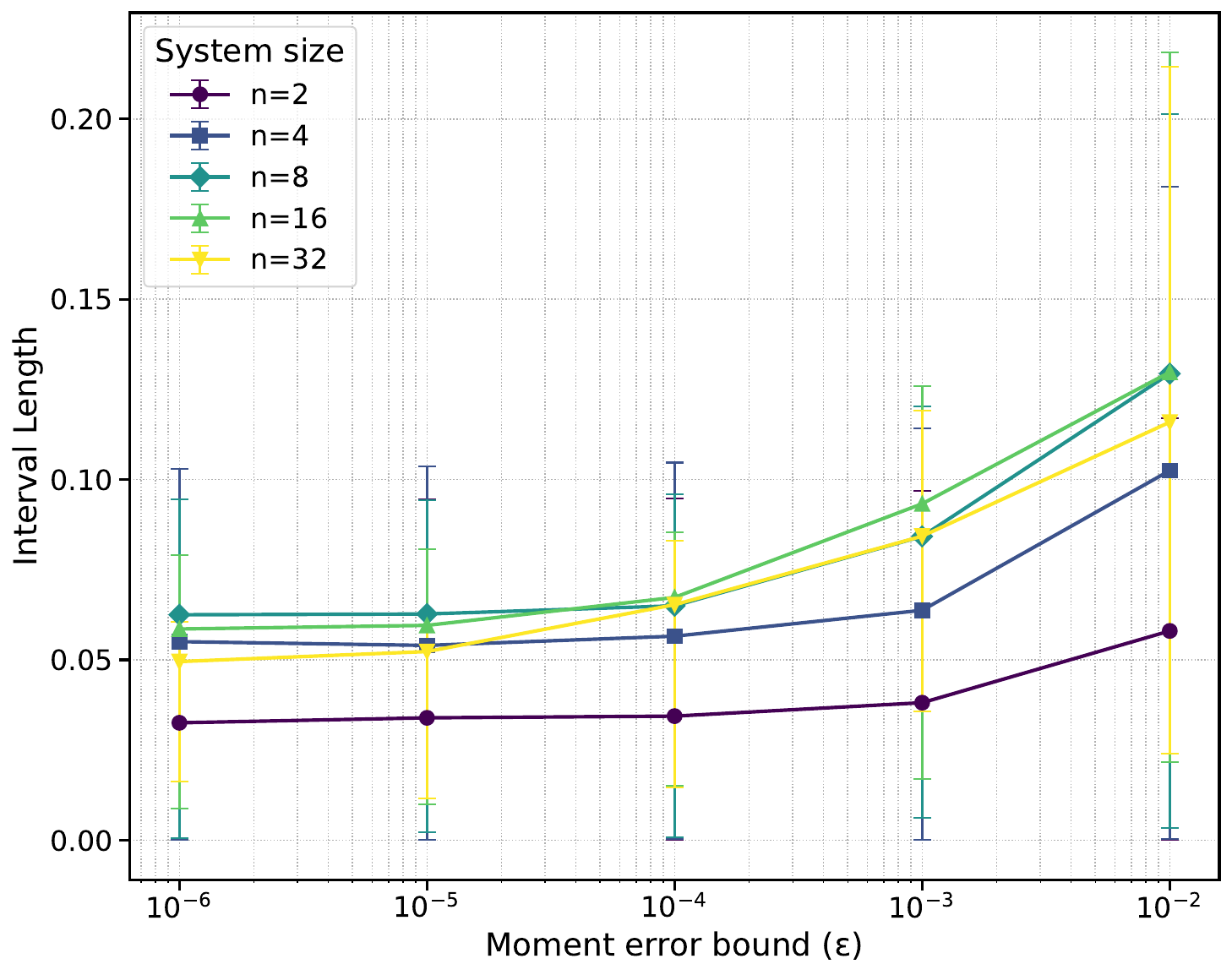}
  \caption{
    The figure shows the mean absolute length of the estimated interval for the maximum eigenvalue ($\lambda_{\max}$) as a function of the moment error bound, $\varepsilon$. Different curves correspond to quantum states of different ranks, $n$. Each data point is the average of 1000 independent trials, where for each trial, a random spectrum was generated from a Dirichlet distribution. The moments up to order $k=4$ were then perturbed by noise bounded by $\varepsilon$, as described in the main text. 
  }
  \label{fig:interval_max_eig}
\end{figure}

The numerical simulations are conducted on one of the servers.
Fig.~\ref{fig:interval_max_eig} summarizes the Monte Carlo data.  
First, the curves are almost straight lines on the log-linear scale, showing that the interval width decreases roughly proportionally to $\varepsilon$ for every rank.
Second, at any fixed $\varepsilon$ the interval widens as the rank increases, reflecting the fact that highly mixed states have flatter spectra.  
The growth, however, is modest: Moving from rank $2$ to rank $32$ increases the mean interval by at most a factor $2$, and the curves start to bunch together for $\varepsilon\le 10^{-4}$, indicating that the upper and lower bounds tighten as soon as the moment estimates become sufficiently accurate.  
Finally, the raw data show that the standard deviation of the interval length never exceeds $0.05$ and decreases for both smaller $\varepsilon$ and larger $n$, which confirms that the estimator is statistically stable across random spectra.

In the most demanding setting we tested ($ n=32$, $ \varepsilon=10^{-3}$), the mean absolute interval width is approximately $0.085$, yielding a small and reliable interval for the largest eigenvalue.
With sampling budgets on the order of $10^{6}$ copies, which is well within reach of current NISQ devices, our moment-based protocol therefore yields practically tight estimates of $\lambda_{\max}$.

While other methods have been proposed for eigenvalue estimation, they often face significant practical hurdles. For instance, the QSF approach~\cite{yao2024nonlinear} requires approximating step functions, a task known to be resource-intensive and sensitive to noise on near-term devices. Another protocol~\cite{pmlr-v247-grier24a} can learn the principal eigenstate to estimate the maximum eigenvalue, but it relies on strong assumptions about the initial state overlap that may not be satisfied in many scenarios. In contrast, our method offers a more direct and practical pathway. By obtaining moments up to order four with an additive error of approximately $10^{-3}$, our framework can constrain the maximum eigenvalue to an interval of length $\sim 0.05$ in most cases. 

Taken together, these findings demonstrate that a few low-order moments are sufficient to effectively localize the maximum eigenvalue of an unknown quantum state. Crucially, these moments are extracted using the resource-optimized circuits presented in Section~\ref{sec:moment_result}, highlighting the efficiency of our approach.


\subsection{Quantum Virtual Cooling}
\begin{figure*}
    \centering
    \includegraphics[width=\linewidth]{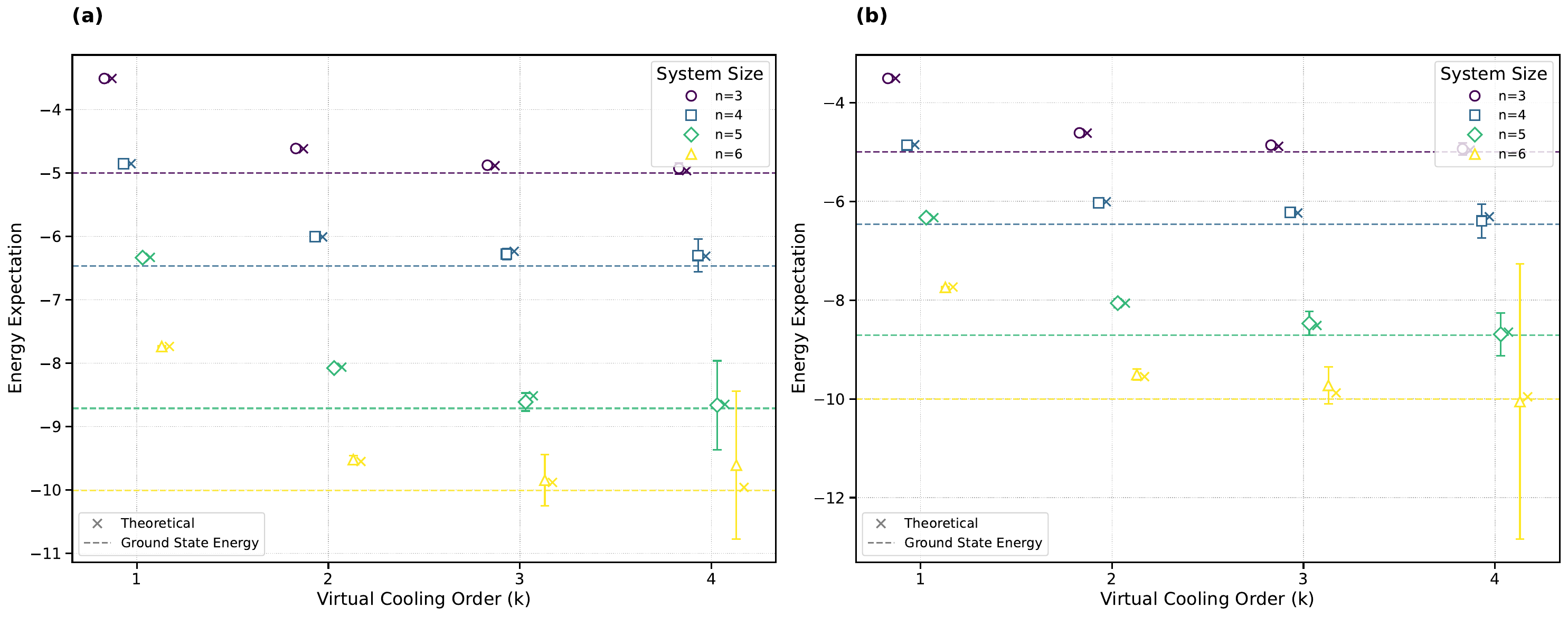}
    \caption{Performance and error analysis for energy estimation of the Heisenberg model, comparing the QVC algorithm against the generalized SWAP test. All simulations were performed with a fixed measurement budget of $10^5$ shots per data point. Simulated values represent the mean of 10 independent runs, with error bars indicating the corresponding standard deviation. Exact theoretical values are marked with 'x' for comparison. Data for different system sizes are distinguished by color and are slightly offset horizontally for clarity. The horizontal dashed lines represent the true ground-state energy for each system size, with colors matching the data points. (a)Simulated energy expectation values obtained using our QVC protocol for different system sizes ($n$) and cooling orders ($k$). (b)Corresponding energy expectation values obtained using the generalized SWAP test under identical simulation conditions. }
    \label{fig:qvc}
\end{figure*}

QVC is a class of protocols that extract low-temperature equilibrium properties from thermal quantum states without physically lowering the bath temperature. This approach enables the study of ground-state and low-energy phenomena using experimentally accessible states and is closely related to virtual distillation and multi-copy error mitigation methods~\cite{william2021Virtual, Vikstal2024studyofnoisein}. 
Practical limitations include the number of copies required, sampling overhead, and sensitivity to noise and imperfect operations.

QVC exploits the identity
\begin{equation}
\label{eq:quantum_virtual_cooling_1}
    \operatorname{Tr}\bigl(O\rho(T/k)\bigr)
= \frac{\operatorname{Tr}\bigl(O\rho(T)^k\bigr)}
       {\operatorname{Tr}\bigl(\rho(T)^k\bigr)},
\end{equation}
where \(\rho(T) = \mathrm{e}^{-\beta H}/\operatorname{Tr}(\mathrm{e}^{-\beta H})\) represents the Gibbs state at temperature \(T\), \(\beta = 1/(k_B T)\), and $H$ denotes the system Hamiltonian.
This enables inference of low-temperature expectation values from nonlinear functionals of the high-temperature Gibbs state $\rho(T)$.
The original protocol requires preparing \(k\) coherent copies of \(\rho(T)\) and measuring suitable swap operators to get unbiased estimators of the numerators and denominators in Eq.~\eqref{eq:quantum_virtual_cooling_1}.
Reconstructing the temperature profile $\bigl\{\operatorname{Tr}\bigl( O\rho(T/k')\bigr)\bigr\}_{k'=2}^k$ via Eq.~\eqref{eq:quantum_virtual_cooling_1} requires estimating the complete sets of numerators $\bigl\{\operatorname{Tr}\bigl( O\rho(T)^{k'}\bigr)\bigr\}_{k'=2}^k$ and denominators $\bigl\{\operatorname{Tr}\bigl( \rho(T)^{k'}\bigr)\bigr\}_{k'=2}^k$. Standard protocols address this by repeating the estimation process for each value of $k'$ independently. Consequently, the total sample complexity to acquire these complete sets of moments accumulates to $\mathcal{O}(k^{2}\log k)$.

In stark contrast, our method, which leverages a single unified circuit, can estimate all of these required numerators and denominators simultaneously. This parallelized approach eliminates the need for repeated, independent experiments, reducing the total sample complexity to just $\mathcal{O}(k\log k)$. This reduction by a linear factor in $k$ represents a significant polynomial speedup, making our approach critically more efficient for the practical implementation of virtual cooling algorithms on near-term quantum hardware.

To demonstrate the performance of this approach, we implement QVC for the one-dimensional Heisenberg model with open boundary conditions,
\begin{equation*}
    H = J \sum_{i=1}^{n-1} (X_iX_{i+1} + Y_iY_{i+1} + Z_iZ_{i+1}) + h \sum_{i=1}^{n} Z_i,
\end{equation*}
focusing on the parameter regime \(J=1\) and \(h=1\). 
Circuits are generated and simulated using Qiskit~\cite{qiskit2024}. 
We consider an initial inverse temperature \(\beta=0.5\), system sizes \(n=3,4,5,6\), and cooling orders (replica numbers) \(k=1,2,3,4\). 
For each $n$, we repeat the simulation multiple times with independent seeds to generate robust statistics for the energy expectation value, allowing for a clear quantification of the standard deviation arising from finite measurement shots.

To provide a practical benchmark of our QVC protocol under realistic experimental constraints, we performed 10 independent simulations for each configuration at a fixed shot budget of $N=10^5$. The results, summarized in Fig~\ref{fig:qvc}(a), demonstrate a systematic and reproducible reduction in energy as the cooling order $k$ increases. For the $n=4$ system, for instance, the mean energy progressively cools from $-4.854\pm0.011$ ($k=1$), to $-6.001\pm 0.043$ ($k=2$), to $-6.282\pm 0.098$ ($k=3$), and finally to $-6.310\pm 0.259$ ($k=4$). This substantially closes the gap to the theoretical ground-state energy of -6.464, with error bars denoting the standard deviation ($\sigma_E$) across independent runs.

A key feature of the QVC protocol is its ability to systematically approach the ground state by increasing the cooling order, $k$. This enhanced cooling, however, is accompanied by increased statistical uncertainty. As expected, since higher $k$ and larger $n$ require more complex collective measurements, sampling noise is amplified under a finite shot budget. This manifests as a growth in both the mean absolute deviation (MAD) from the exact energy and the run-to-run standard deviation ($\sigma_E$). For instance, at the highest cooling order presented ($k=4$), the MAD grows from 0.081 at $n=3$ to 0.941 at $n=6$, and the corresponding $\sigma_E$ surges from $\approx 0.088$ to $\approx 1.139$.

Crucially, this increased variance does not imply a loss of accuracy in the mean. The protocol exhibits negligible systematic bias across the configurations we studied. In the $n=5, k=2$ case, for example, the mean simulated energy ($-8.065$) deviates from the exact value ($-8.062$) by merely $\approx 0.004$, an amount less than a tenth of the statistical uncertainty ($\sigma_E \approx 0.054$). This confirms that the observed variance is predominantly statistical in nature, and the protocol remains robustly centered on the true value.

Having established the statistical properties of our protocol, we rigorously benchmarked its resource efficiency against the generalized SWAP test, a standard method for measuring expectation values of the form $\operatorname{Tr}(H\rho^k)$. To ensure a fair assessment of resource utilization, the comparison was performed under a fixed budget of prepared state copies. The results reveal a pronounced and systematic advantage of our QVC protocol.

The most significant disparity emerges in the statistical noise. For an identical resource cost, the generalized SWAP test consistently exhibits substantially higher variance, a deficiency that is exacerbated for larger systems and higher cooling orders. For instance, in the challenging $n=6, k=4$ configuration, the SWAP test yields a standard deviation of $\sigma_E \approx 2.57$. This is more than a factor of two larger than the $\sigma_E \approx 1.14$ observed for our QVC protocol. This implies that achieving a comparable level of statistical precision with the SWAP test would necessitate an approximately fivefold increase in the number of measurements, highlighting a quadratic scaling disadvantage in resource cost.

This inferior precision directly translates to a degradation in reliability. For the $n=6, k=4$ case, the MAD for the SWAP test is $\approx 2.13$, again more than double the value of $\approx 0.94$ achieved with our method. While the mean estimated energies of the two methods are more comparable in less demanding configurations (e.g., for $n=4, k=3$), the significantly higher variance of the SWAP test renders it fundamentally less reliable. A large standard deviation implies that any single experimental realization is substantially more likely to yield an estimate that deviates significantly from the true value, undermining its practical utility.

In summary, the QVC protocol demonstrates a superior performance profile for an equivalent resource budget. It yields estimates that are not only more precise (lower variance) but also more consistently accurate, particularly in the demanding regime of large system sizes and high cooling orders. This highlights the resource-efficiency of our proposed method, establishing it as a more viable and scalable approach for ground- and thermal-state characterization on near-term quantum hardware.


To conclusively validate our protocol's feasibility on contemporary quantum hardware, we performed a proof-of-principle experiment to measure the integer R\'{e}nyi entropy on a real superconducting quantum processor. 
This experiment serves as a direct demonstration of our method's applicability in a noisy, real-world environment. 
The detailed experimental setup, procedure, and data analysis are presented in Appendix~\ref{appendix_B}.

\section{Conclusions and discussions}

In this work, we have developed and experimentally validated a highly efficient protocol for estimating nonlinear functionals of a quantum state, addressing a central challenge in quantum system characterization and property spectroscopy. Our contributions are multifaceted and establish a new state-of-the-art in resource efficiency and practical applicability. First, our protocol provides a unified framework for simultaneously estimating the entire hierarchy of moments, $\text{Tr}(\rho^j)$ for $j=2, \dots, k$, as well as general polynomial functionals $f(\rho)$ and their observable-weighted counterparts $f(O, \rho)$. This versatility is achieved without incurring significant resource overhead. The circuit depth scales efficiently with the moment order, and the qubit requirement is kept modest, in contrast to competing methods. This efficiency is further enhanced by an adaptive measurement scheme with qubit reset, which allows for the sequential reuse of auxiliary qubits to minimize the overall qubit footprint. Critically, our protocol also attains a near-optimal sample complexity of $\mathcal{O}(k \log k / \varepsilon^2)$. This constitutes a quadratic enhancement over naive sequential strategies and underscores the fundamental efficiency of our coherent approach.

To demonstrate the protocol's practical viability and robustness for near-term quantum hardware, We successfully measured the R\`{e}nyi entropy from the second to the fourth order on the 133-qubit \verb|ibm_torino|~\cite{ibm_torino} superconducting processor. The experimental results, once subjected to standard error mitigation, align well with theoretical predictions, confirming our method's robustness and its immediate potential for deployment on near-term quantum devices.

Beyond its direct applications, our framework resolves a key problem in quantum property estimation and provides a new avenue to design qubit-efficient quantum algorithms for quantum functionals estimation via the CSWAP-based data reuploading technique. Specifically, it extends naturally to the simultaneous estimation of $\mathrm{Tr}(\rho\sigma)$, $\mathrm{Tr}\bigl((\rho\sigma)^2\bigr)$,$\dots$, $\mathrm{Tr}\bigl((\rho\sigma)^k\bigr)$ with the same sample complexity, thereby answering an open problem posed in~\cite{chen2025simultaneous}. This capability paves the way for novel quantum algorithms targeting critical multi-state quantities like trace distance and relative entropy. More broadly, the efficient measurement of higher-order moments provides a powerful new primitive for applications ranging from entanglement spectroscopy and quantum machine learning to the verification of complex quantum simulations. Looking forward, a practical avenue is to integrate our protocol with advanced error mitigation techniques for enhanced performance on noisy hardware. Investigating hardware-aware optimizations of the CSWAP circuits also remains an exciting direction. On the theoretical front, extending these methods to estimate $\mathrm{Tr}(\rho^j)$ and $\mathrm{Tr}(O\rho^j)$ for non-integer $j$ presents an interesting direction.
\begin{acknowledgments}
We thank Lei Zhang, Zhan Yu, and Yingjian Liu for helpful discussions. This work was partially supported by the National Key R\&D Program of China (Grant No.~2024YFB4504004), the National Natural Science Foundation of China (Grant. No.~12447107), the Guangdong Provincial Quantum Science Strategic Initiative (Grant No.~GDZX2403008, GDZX2403001), the Guangdong Provincial Key Lab of Integrated Communication, Sensing and Computation for Ubiquitous Internet of Things (Grant No.~2023B1212010007), the Quantum Science Center of Guangdong-Hong Kong-Macao Greater Bay Area, and the Education Bureau of Guangzhou Municipality.
\end{acknowledgments}

\bibliography{ref}


\newpage

\appendix
\setcounter{subsection}{0}
\setcounter{table}{0}
\setcounter{figure}{0}

\vspace{3cm}
\onecolumngrid
\vspace{2cm}

\begin{center}
\Large{\textbf{Appendix for Near-Optimal and Resource-Efficient Simultaneous Estimation of Nonlinear Quantum State Functionals} \\ \textbf{
}}
\end{center}

\renewcommand{\theequation}{S\arabic{equation}}
\renewcommand{\theproposition}{S\arabic{proposition}}
\renewcommand{\thedefinition}{S\arabic{definition}}
\renewcommand{\thefigure}{S\arabic{figure}}
\setcounter{equation}{0}
\setcounter{table}{0}
\setcounter{section}{0}
\setcounter{proposition}{0}
\setcounter{definition}{0}
\setcounter{figure}{0}


\section{Details of proof techniques}\label{appendix_A}

\subsection{Proof of Theorem~\ref{thm:theorem_1}}\label{appendix:proof_of_theorem_1}

\renewcommand\theproposition{\ref{thm:theorem_1}}
\begin{theorem}
Let $\rho$ be an arbitrary $m$-qubit quantum state. 
For any positive integer $k$, all moments up to order $k$, $\operatorname{Tr}(\rho^{2}), \dots, \operatorname{Tr}(\rho^{k})$, can be estimated simultaneously using a circuit on $2m + 1$ qubits. 
This estimation achieves an additive error $\varepsilon$ with probability at least $2/3$, using $\mathcal{O}\bigl(k \log k/ \varepsilon^{2}\bigr)$ copies of $\rho$.
\end{theorem}
\renewcommand{\theproposition}{S\arabic{proposition}}

\begin{proof}
Our proof focuses on finding a list of commute observables and a practical circuit that can be used to estimate all these moments up to order $k$, unlike previous approaches that employ different circuits for each moment.


Consider a circuit in which we employ \(k-1\) auxiliary qubits \(A_1, \dots, A_{k-1}\) initialized to \(\ket{0}\), along with \(k\) identical copies of an unknown mixed state \(\rho\) stored in registers \(B_1, \dots, B_k\).  
A Hadamard gate prepares each auxiliary qubit in the uniform superposition state \(\ket{+}\).
Each auxiliary qubit then controls a CSWAP (Fredkin) gate that swaps \(B_1\) with another \(\rho\)-register: \(A_1\) controls SWAP(\(B_1, B_2\)), \(A_2\) controls SWAP(\(B_1, B_3\)), and so on through \(A_{k-1}\) controlling SWAP(\(B_1, B_k\)).
At the end of the circuit, we projectively measure the auxiliary qubits in the \(X\)-basis.

By measuring each auxiliary qubit immediately after its CSWAP gate, then resetting and reusing these qubits while similarly resetting and reusing the $\rho$-registers $B_2$ through $B_k$ post-SWAP, we obtain the circuit configuration shown in Fig.~\ref{fig:theorem1_moment}.

We now demonstrate that these measurements suffice to capture the resulting statistics. 
We introduce the family of observables
\begin{equation*}
\begin{aligned}
\hat{O}_1 &= I_{A_1} \otimes \cdots \otimes I_{A_{k-2}}\otimes X_{A_{k-1}} \otimes I_B, \\
\hat{O}_2 &= I_{A_1} \otimes \cdots \otimes I_{A_{k-3}}\otimes X_{A_{k-2}}\otimes X_{A_{k-1}} \otimes I_B, \\
&\;\; \vdots \\
\hat{O}_{k-1} &= X_{A_1} \otimes \cdots \otimes X_{A_{k-1}} \otimes I_B,
\end{aligned}
\end{equation*}
where $I_{A_i}$ denotes the identity on the $i$-th auxiliary qubit and $I_B$ acts on the Hilbert space of the $k$ copies of $\rho$.

Let \(S_{ij}\) denote the unitary operator that swaps the \(i\)-th and \(j\)-th tensor factors of a state.
We first evaluate the expectation value of the observable \(\hat O_{1}\).  
After the circuit, \(\hat O_{1}\) projects the auxiliary register onto the two eigenvalues of a single Pauli-\(X\), and averaging over these outcomes produces
\[
\langle\hat O_{1}\rangle
   =\tfrac12\operatorname{Tr}\bigl(S_{1,k}\rho^{\otimes k}\bigr)
   +\tfrac12\operatorname{Tr}\bigl(\rho^{\otimes k}S_{1,k}\bigr)
   =\operatorname{Tr}\bigl(\rho^{2}\bigr),
\]
where the last equality follows from the cyclicity of the trace.

The construction for \(\hat O_{2}\) is analogous but involves two auxiliary qubits. 
Here, two auxiliary qubits are measured in the \(X\)-basis, so four equally likely sign combinations contribute:
\[
\begin{aligned}
\langle\hat O_{2}\rangle &= \tfrac14\mathrm{Tr}(S_{1,k}S_{1,k-1}\rho^{\otimes k})+
\tfrac14\mathrm{Tr}(S_{1,k}\rho^{\otimes k}S_{1,k-1})+
\tfrac14\mathrm{Tr}(S_{1,k-1}\rho^{\otimes k}S_{1,k})+
\tfrac14\mathrm{Tr}(\rho^{\otimes k}S_{1,k-1}S_{1,k}).
\end{aligned}
\]
By the cyclic property of the trace, each of the four terms evaluates to $\operatorname{Tr}(\rho^3)/4$, so the expectation value simplifies to $\langle\hat O_{2}\rangle=\operatorname{Tr}(\rho^{3})$.

The preceding two cases make the general pattern transparent. For general \(1\le j\le k-1\), the non-identity factors in \(\hat O_{j}\) are products of \(S_{1,k-j+1}\), $\dots$, \(S_{1k}\). 
Any ordering of these \(j\) operators yields the same \((j+1)\)-cycle, which we denote \(P_{j+1}\).  
Because there are \(2^{j}\) possible outcome strings, each weighted by \(2^{-j}\), the expectation value simplifies to
\begin{equation*}
\begin{aligned}
\langle\hat O_{j}\rangle &=2^{j}\cdot2^{-j}
     \operatorname{Tr}\bigl(P_{j+1}\rho^{\otimes k}\bigr)=\operatorname{Tr}\bigl(\rho^{j+1}\bigr).
\end{aligned}
\end{equation*}

Therefore, we can get the expectation of \(\hat O_{1}\), $\hat O_{2}$,~$\dotsc$, \(\hat O_{{k-1}}\) as follows:
\[
\mathbb{E}[\hat O_{1}]=\operatorname{Tr}(\rho^{2}),\quad
\ldots ,\quad
\mathbb{E}[\hat O_{k-1}]=\operatorname{Tr}(\rho^{k}).
\]

Running the circuit once produces a string of measurement outcomes 
\((x_1,x_2,\dots ,x_{k-1})\in\{\pm1\}^{k-1}\),  
where \(x_i\) is the eigenvalue obtained from the \(X\)-basis read-out of \(A_i\).  
Because every operator \(\hat O_m\) is a tensor product of either \(X\) or \(I\) on the auxiliary register, all \(\hat O_m\) commute and share the global \(X\)-basis as an eigenbasis.
Hence, a single shot already fixes the eigenvalue of every \(\hat O_m\):
\[
\omega_m(x_1,\dots ,x_{k-1})=\prod_{i=k-m}^{k-1}x_i, \quad m=1,\dots ,k-1.
\]

In other words, one merely multiplies the outcomes of those qubits on which \(\hat O_m\) contains an \(X\); qubits accompanied by an identity factor do not affect the product.  Repeating the experiment \(N\) times and averaging the resulting \(\omega_m\) values,
\begin{equation*}
    \begin{aligned}
        \langle\hat O_m\rangle \approx\frac{1}{N}\sum_{\ell=1}^{N}\omega_m^{(\ell)},
    \end{aligned}
\end{equation*}
yields an unbiased estimator of each expectation value \(\langle\hat O_m\rangle\) without the need for additional circuits or measurements.
Thus, the entire set \(\{\langle\hat O_1\rangle\), $\dots$, \(\langle\hat O_{k-1}\rangle\}\) is extracted simultaneously from the same measurement record, making the procedure experimentally efficient.

Now, let us turn to take a look at the sample complexity.
Given $kn$ copies of $\rho$, we execute the circuit $n$ times.
Denote the outcome of $\hat{O}_j$ in $i$-th run as $\mu_{j, i}$.
Note that $\hat{O}_j^2 = I$, and $\mu_{j, i} \in \{-1, 1\}$, we have $\mathbb{E}[\mu_{j, i}]=\langle \hat{O}_j \rangle = \mathrm{Tr}(\rho^{j+1})$, and $\mathrm{Var}[\mu_{j, i}] = \langle \hat{O}_j^2 \rangle - \langle \hat{O}_j \rangle^2=1-\langle \hat{O}_j \rangle^2\leq 1$.

For $j=1, \dotsc, k-1$, applying Hoeffding’s inequality for any $t>0$, we have
\begin{equation*}
\begin{aligned}
    \mathbb{P}\Biggl(\Biggl\lvert \sum_{i=1}^n \mu_{j, i} -n\mathrm{Tr}(\rho^{j+1})\Biggr\rvert \geq t\Biggr) \leq 2 \exp \bigl( -2t^2 / (4n) \bigr).
\end{aligned}
\end{equation*}
Taking $\overline{\mu}_j=\frac{1}{n}\sum_{i=1}^n \mu_{j, i}$ and $t = n\varepsilon$, we have
\begin{equation*}
    \begin{aligned}
        \mathbb{P}\bigl(\lvert \overline{\mu}_j- \mathrm{Tr}(\rho^{j+1})\rvert \geq \varepsilon \bigr)
\leq 2\exp (-n\varepsilon^2 / 2).
    \end{aligned}
\end{equation*}

We aim to bound the total probability of failure for all $k-1$ estimators by a constant, conventionally chosen as $1/3$. By applying the union bound, this requires the failure probability of each individual estimator to be bounded by $1/(3k)$. Combining this with Hoeffding's inequality leads to the choice of the number of measurements $n$.

Taking $n = \lceil 2 \log(6k)/\varepsilon^2 \rceil$, we have $\mathbb{P}\bigl(\lvert \overline{\mu}_j- \mathrm{Tr}(\rho^{j+1})\rvert \geq \varepsilon \bigr)\leq 1/(3k)$ for each $j \in \{1, \dots, k-1\}$.

By the union bound, the probability that at least one of the $k-1$ estimators fails is bounded by:
\begin{equation*}
\begin{aligned}
\mathbb{P}\left( \bigcup_{j=1}^{k-1} \{ \lvert \overline{\mu}j- \mathrm{Tr}(\rho^{j+1})\rvert \geq \varepsilon \} \right)
&\leq \sum_{j=1}^{k-1}\mathbb{P}\bigl(\lvert \overline{\mu}j- \mathrm{Tr}(\rho^{j+1})\rvert \geq \varepsilon\bigr) \\
&\leq \sum_{j=1}^{k-1} \frac{1}{3k} = \frac{k-1}{3k} \leq \frac{1}{3}.
\end{aligned}
\end{equation*}

Total sample complexity as $\mathcal{O}\bigl(k\log k/\varepsilon^2\bigr)$ ensures that all $k-1$ estimates succeed with probability at least $2/3$, which completes the proof.

\end{proof}
\subsection{Proof of Theorem~\ref{thm:theorem_2}}\label{appendix:proof_of_theorem_2}

\renewcommand\theproposition{\ref{thm:theorem_2}}
\begin{theorem} 
Let $\rho\in\mathbb{C}^{2^{m}\times 2^{m}}$ be an arbitrary $m$-qubit state and let $f(\rho)$ be the degree-$k$ polynomial state functional defined previously.
For any $\varepsilon>0$, $f(\rho)$ can be estimated using a circuit on $2m+\lceil\log_{2}\! k\rceil+1$ qubits within additive error $\varepsilon$ with success probability at least $2/3$, using $\mathcal{O}\bigl(k\|f\|_{1}^{2}/\varepsilon^{2}\bigr)$ copies of $\rho$.
The circuit employs $\mathcal{O}(k)$ CSWAP gates, $\mathcal{O}(k)$ qubit-reset operations, and requires only single-qubit measurements.
\end{theorem}
\renewcommand{\theproposition}{S\arabic{proposition}}
\begin{proof}

Let $\rho$ be a density operator on a $d$-dimensional Hilbert space $\mathcal H$.  Throughout this section, we consider the joint Hilbert space  
\begin{equation*}
    \mathbb{C}^{2}_{A}\;\otimes\;\bigl(\mathbb{C}^{2}\bigr)^{\otimes\lceil\log_{2}\! k\rceil}_{B}\;\otimes\;\mathcal{H}^{\otimes k},
\end{equation*}
which comprises a single-qubit control register $A$, an $\lceil\log_{2}\! k\rceil$-qubit register $B$ that stores the binary representation of an index $j\in\{0,1,\dots ,k-1\}$, and $k$ identical subsystems each prepared in the state $\rho$. 

To estimate $f(\rho)$, we first rewrite the function by introducing the normalised coefficients $\lambda_{j}:=\alpha_{j}/\|f\|_{1}$ for $j=1,\dots ,k$, where $\|f\|_{1}$ is $\ell_1$-norm of the coefficient vector. This allows us to express $f(\rho)$ as:
\begin{equation*}
    \begin{aligned}
    f(\rho)=\|f\|_{1}\sum_{j=1}^{k}\lambda_{j}\operatorname{Tr}\bigl(\rho^{j}\bigr),
    \end{aligned}
\end{equation*}
By definition, these normalized coefficients satisfy $\sum_{j=1}^{k}\lvert\lambda_{j}\rvert=1$. This normalization separates the overall magnitude \(\|f\|_{1}\) from the direction specified by the vector \((\lambda_{1},\dots,\lambda_{k})\), which simplifies both the implementation and analysis of the associated quantum algorithm.

To simplify the initial presentation, assume $\lambda_1$,~$\dotsc$, $\lambda_k$ are non-negative.
We first apply a Hadamard gate to the first auxiliary qubit, transforming it into the $\ket{+}$ state.
The algorithm then interleaves a series of rotations on System $B$ with controlled-SWAP operations. The rotations, denoted $V_j$ for $j = 1$,~$\dotsc$, $k-1$, are designed to incrementally prepare the desired amplitudes. To this end, we define the cumulative weights as $\Lambda_j := \sum_{i=j}^k \lambda_i.$ For $1 \leq j \leq k-1$, we act on the two-dimensional subspace $\mathrm{span}\{ \ket{j-1}, \ket{j} \}$ with the Givens rotation
\begin{equation*}
    \begin{aligned}
    V_j \;=\; \bigl[G(c_j)\bigr]_{\{|j-1\rangle,|j\rangle\}} \oplus \mathbb{I}_{\text{rest}},
\end{aligned}
\end{equation*}
where $G(c_j)= \begin{pmatrix} c_j & s_j\\[2pt] -s_j & c_j \end{pmatrix}$, $c_j=\sqrt{\lambda_j/\Lambda_j}$, $s_j=\sqrt{\Lambda_{j+1}/\Lambda_j}$.
Since $G(c_j)\in\mathrm{SO}(2)$, each $V_j$ is unitary. 
A straightforward induction shows that if these rotations were applied in sequence, they would produce the target state $|\lambda\rangle$:
\begin{equation*}
    \begin{aligned}
    V_{k-1}\cdots V_1\ket{0} = \sum_{i=1}^k \sqrt{\lambda_i}\ket{i-1} \equiv |\lambda\rangle.
\end{aligned}
\end{equation*}

Our circuit, however, implements these steps in an interleaved fashion, as we now describe.

First, we apply the unitary \(V_1\) on System \(B\). 
As described above, this operation transforms \(\ket{0}\) to  
\[
\sqrt{\lambda_1} \ket{0} + \sqrt{\Lambda_2}\ket{1},
\]
which is readily implemented by an \(R_y\) rotation on the last qubit.

Subsequently, we perform a CSWAP operation between the first two copies of \(\rho\), conditioned jointly on System \(A\) being in \(\ket{1}\) and System \(B\) being in state \(\ket{1}\). 
The operator for this controlled operation is given by
\[
\ket{1}\bra{1}_A \otimes \ket{1}\bra{1}_B \otimes \mathtt{SWAP}_{12}.
\]

Following this, we execute the unitary \(V_2\) on System \(B\). 
Notably, \(V_2\) leaves the \(\ket{0}\) state unchanged, i.e., \(V_2\ket{0}=\ket{0}\), and rotates the \(\ket{1}\) state as
\[
V_2\ket{1} = \sqrt{\frac{\lambda_2}{\Lambda_2}}\ket{1} + \sqrt{\frac{\Lambda_3}{\Lambda_2}}\ket{2}.
\]
Thereafter, we perform another CSWAP between the first and third copies of \(\rho\), which is activated when System \(A\) is in \(\ket{1}\) and System \(B\) is in \(\ket{2}\). 
Its density matrix representation is
\[
\ket{1}\bra{1}_A \otimes \ket{2}\bra{2}_B \otimes \mathtt{SWAP}_{13}.
\]

It is important to note that after the first CSWAP, System \(B\) evolves into a superposition of \(\ket{0}\), \(\ket{1}\), and \(\ket{2}\). 
However, because the initial CSWAP was conditioned on \(B\) being in \(\ket{1}\), only the component corresponding to \(\ket{1}\) undergoes that SWAP. 
Since $V_2$ preserves the \(\ket{0}\) state while rotating \(\ket{1}\) into a superposition that includes \(\ket{2}\), after applying \(V_2\), the branch in which System \(B\) is in \(\ket{2}\) controls the subsequent SWAP between the first and third copies of \(\rho\).

Thus, for System \(B\), the overall action is as follows:
- In the \(\ket{0}\) branch, no SWAP is executed.
- In the \(\ket{1}\) branch, the SWAP operation between copies 1 and 2 is enacted.
- In the \(\ket{2}\) branch, both the SWAP between copies 1 and 2 and the SWAP between copies 1 and 3 are performed.

This procedure is extended by consecutively applying \(V_j\) on System \(B\) (where \(j = 1,2,\ldots\)) and, after each \(V_j\), executing a SWAP when System \(A\) is in \(\ket{1}\) and System \(B\) is in the state \(\ket{j}\). 
In our circuit implementation, the \(V_j\) unitaries can be realized as a controlled-$R_y$ gate via a Gray code decomposition. 

Note that each $V_j$ is a Givens rotation and takes the form
\[
V_j = \begin{bmatrix}
    I_1 & & \\
    & U_j & \\
    & & I_2
\end{bmatrix},
\]
where $I_1$ denotes the $(j-1) \times (j-1)$ identity matrix, $I_2$ is an identity matrix of appropriate dimension, and $U_j$ is a $2 \times 2$ unitary matrix. 
Under the Gray code encoding, adjacent integers $j-1$ and $j$ differ by exactly one bit in their binary representations. 
Consequently, $V_j$ can be implemented as a controlled-$R_y$ gate. 
Specifically, suppose the binary representations of $j-1$ and $j$ under the Gray code encoding are $i_1 \cdots i_{t-1} i_t i_{t+1} \cdots i_m$ and $i_1 \cdots i_{t-1} i'_t i_{t+1} \cdots i_m$, respectively, where $m = \lceil \log_2\! k \rceil$ and $i_t \neq i'_t$. 
Then $V_j$ corresponds precisely to a controlled-$R_y$ gate applied to the $t$-th qubit, conditioned on the values $i_1, \dotsc, i_{t-1}, i_{t+1}, \dotsc, i_{m}$ of all other qubits. 
The rotation angle $\theta$ for this controlled-$R_y$ gate is given by
\[
\theta = 2 \arccos \Biggl( \sqrt{\frac{\lambda_j}{\Lambda_j}} \Biggr). 
\]
And this case is shown in Fig.~\ref{fig:graycodeCRy}.

The case where $i_t = 1$ and $i'_t = 0$ requires no special consideration. Since the measurement outcome probabilities depend only on the magnitude of the state vector components, the global phase introduced in this scenario is physically inconsequential. 
\begin{figure}[htbp]
    \centering
    \begin{quantikz}[transparent]
        \qw^{B} & \gate{V_j} &\\
    \end{quantikz}\quad =
    \begin{quantikz}[transparent]
        \qw^{B_1} & &\ctrl[wire style={"\ket{i_1}"}]{1}&\\
        \qw^{B_2\;\text{to}\;B_{t-1}}&\qwbundle{}&\ctrl[wire style={"\ket{i_2 \ldots i_{t-1}}"}]{1}& \\
        \qw^{B_t} && \gate{R_y(\theta)}&\\
        \qw^{B_{t+1}\;\text{to}\;B_{m-1}}&\qwbundle{}&\ctrl[wire style={"\ket{i_{t+1} \ldots i_{m-1}}"{right}}]{-1}& \\
        \qw^{B_m} && \ctrl[wire style={"\ket{i_{m}}"{right}}]{-1}&
    \end{quantikz}
    \caption{Implementation of $V_j$ assuming the binary representations of $j-1$ and $j$ under the Gray code encoding differ by $t$-th bit.}
    \label{fig:graycodeCRy}
\end{figure}

For the general case where coefficients $\lambda_j$ can be negative, we must encode their signs as relative phases in the superposition. This is achieved by applying a conditional phase flip to the ancilla qubit in System $A$.
Specifically, whenever a coefficient $\lambda_j$ is negative, we apply a CZ gate on System $A$ controlled by the projector $\ketbra{j}{j}$ on System $B$. 
To systematically handle the negative coefficients, we define a sign function for each coefficient as follows:
\begin{equation*}
    \begin{aligned}
    \mathrm{sgn}(\lambda_i)=  \begin{cases}  +1, & \lambda_i > 0,\\[1mm] 0, & \lambda_i = 0,\\[1mm] -1, & \lambda_i < 0. \end{cases}
    \end{aligned}
\end{equation*}

If $\sum_i \mathrm{sgn}(\lambda_i) < 0,$ which indicates that the overall negative contribution dominates, it is advantageous to compute $-f(\rho)$ instead of $f(\rho)$, as this choice reduces the number of negative coefficients that require correction via CZ gates. 
Under typical assumptions about the distribution of signs, this approach requires at most $\lfloor k/2 \rfloor$ CZ gates. 
Subsequently, we perform an $X$-basis measurement on the first auxiliary qubit. 
Similarly, we can leverage reset techniques to reduce the number of qubits. 
Notably, the \(k\) copies of \(\rho\) can be implemented using only \(2m\) qubits, so the total number of qubits required for our circuit is $2m + \lceil \log_2\!k \rceil + 1$. 
The resulting circuit is depicted in Fig.~\ref{fig:QSF_new_circuit}.

Thus, we obtain the expectation value
\[
\langle X \rangle = \sum_{i=1}^k \lambda_i\operatorname{Tr}(\rho^i).
\]
By invoking Hoeffding’s inequality, to achieve an additive error \(\varepsilon/\lVert f\rVert_1\) with a confidence level of at least \(1-\delta\), one must repeat the experiment \(\mathcal{O}\bigl(\lVert f\rVert_1^2\log\delta^{-1} /\varepsilon^2\bigr)\) times. 
Since each experimental run requires the preparation of \(k\) copies of \(\rho\), the overall sample complexity scales as $\mathcal{O}\bigl(k\lVert f\rVert_1^2\log\delta^{-1} /\varepsilon^2\bigr)$. Neglecting the constant dependence on \(\delta\) yields the final complexity of \(\mathcal{O}(k\lVert f\rVert_1^2/\varepsilon^2)\), which completes the proof.

\end{proof}

\subsection{Proof of Corollary~\ref{cor:multi-funcs}} \label{appendix:proof_of_theorem_3}

\renewcommand\theproposition{\ref{cor:multi-funcs}}
\begin{corollary}
Let $\rho$ be an $m$-qubit state. 
Consider a set of $n_f$ polynomial state functionals, $\{f_i(\rho)\}_{i=1}^{n_f}$. 
Each functional $f_i$ has its own degree $k_i$ and coefficients $\alpha_{i,j}$. Let $k:= \max_{1\leq i \leq n_f}k_i$ be the maximum degree in the set.
For any $\varepsilon>0$, the values $f_1(\rho),\dots,f_{n_f}(\rho)$ can be estimated within an additive error $\varepsilon$ with probability at least $2/3$ using $\mathcal{O}\Bigl(k\max_{1\le i\le n_f} ||f_i||_1^{2}\log(\min\{k,n_f\})/\varepsilon^{2} \Bigr)$ copies of $\rho$. 
\end{corollary}
\renewcommand{\theproposition}{S\arabic{proposition}}
\begin{proof}
We first consider the case when $k \leq n_f$.
Applying Theorem~\ref{thm:theorem_1}, we obtain estimates for all moments up to order $k$, each accurate to within an additive error of $\varepsilon / \max_j \lVert f_j \rVert_1$. 
These estimates directly yield approximations for the state functions $f_1(\rho), \dotsc, f_{n_f}(\rho)$, each with an additive error at most $\varepsilon$. 
The sample complexity required for this procedure scales as $\mathcal{O}\bigl(k \log k \max_j \lVert f_j \rVert_1^2 / \varepsilon^2 \bigr)$. 

Next, we consider the case when $k > n_f$.
We implement the circuit shown in Fig.~\ref{fig:multiple-function}, where $V_i^{(j)}$ denotes the unitary $V_i$ (defined in Appendix~\ref{appendix:proof_of_theorem_1}) for the $j$-th function. Consider the observables
\[
\hat{O}_1 = X_{A_1}, \quad \dots, \quad \hat{O}_{n_f} = X_{A_{n_f}}.
\]
Since $\hat{O}_1$ acts as identity on all systems except $A_1$, we need only consider diagonal terms in other systems ($A_2, \dots, B_1, \dots$). 
For any CSWAP gate not controlled by $A_1$, this gate will not affect the outcome of $\hat{O}_1$.
For example, the operation controlled by $A_2$ and $B_2$ preserves the state:
\[
|c_0|^2 S_{1,2} \rho^{\otimes 2} S_{1,2} + |c_1|^2 \rho^{\otimes 2} = \rho^{\otimes 2}
\]
where $c_0, c_1 \in \mathbb{C}$ satisfy $|c_0|^2 + |c_1|^2 = 1$. 
This demonstrates that such gates do not affect the outcome of $\hat{O}_1$. 
By Theorem~\ref{alg:theorem2_singleqsf}, we conclude
\[
\langle \hat{O}_1 \rangle = f_1(\rho), \quad \dots, \quad \langle \hat{O}_{n_f} \rangle = f_{n_f}(\rho).
\]

Executing the circuit $\mathcal{O}\bigl( \log n_f \max_j \lVert f_j \rVert_1^2 / \varepsilon^2 \bigr)$ times yields, for each state functional $f_j(\rho)$, an estimate accurate to within an additive error $\varepsilon$, succeeding with probability at least $1 - 1/(3n_f)$. 
Applying the union bound over the $n_f$ functions guarantees that, with probability at least $2/3$, all state functions are estimated simultaneously within an additive error $\varepsilon$, which completes the proof.
\end{proof}
\begin{figure*}[ht]
    \centering
    \includegraphics[width=0.75\linewidth]{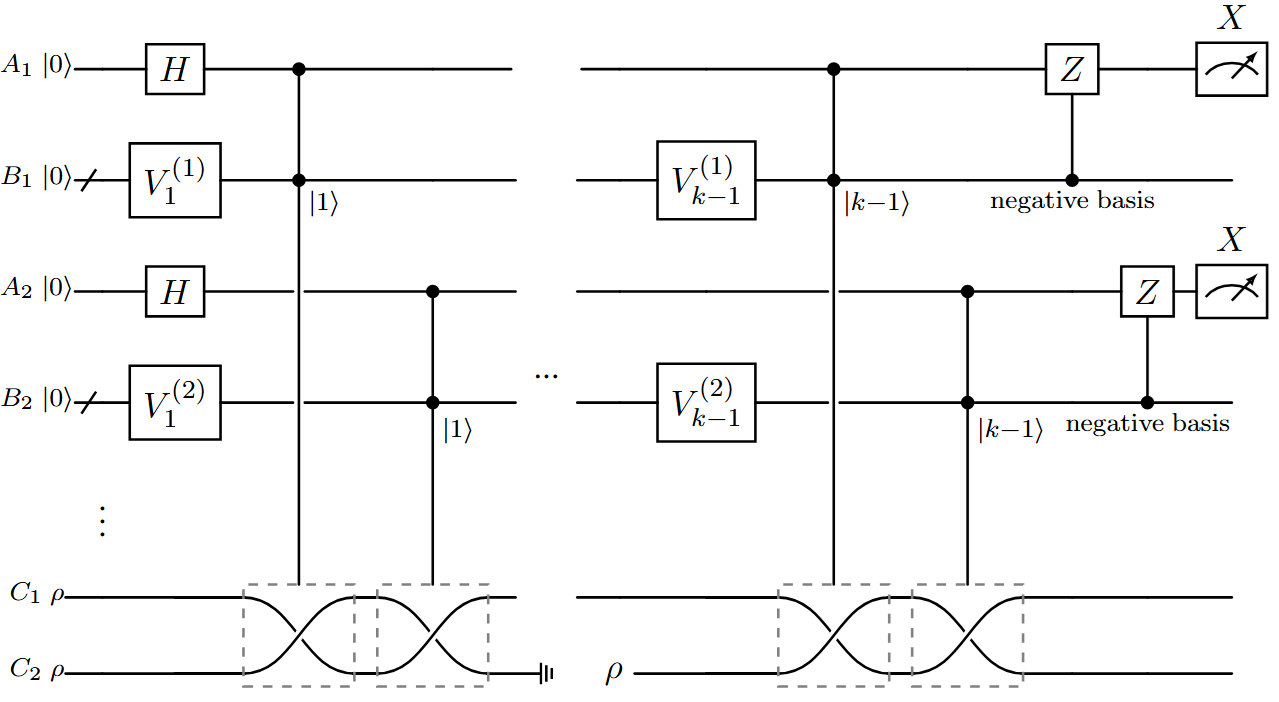}
    \caption{Circuit for multiple state functions estimation.}
    \label{fig:multiple-function}
\end{figure*}


\section{Additional Proofs and Experimental Details}
\subsection{Derivation of Bounds for the Maximum Eigenvalue}
\label{appendix:proof_of_interval}

We derive the upper and lower bounds on the maximum eigenvalue, $\lambda_{\max}$, of a density operator $\rho$. The true $j$-th moment is $m_j = \mathrm{Tr}(\rho^j) = \sum_i \lambda_i^j$, and its unbiased estimate $\hat{m}_j$ is subject to a bounded error $|\hat{m}_j - m_j| \le \varepsilon$. This implies $m_j \in [ \hat{m}_j - \varepsilon, \hat{m}_j + \varepsilon ]$.

The derivation of the upper bound starts from the definition of the moment $m_j$:
\begin{equation*}
m_j = \sum_i \lambda_i^j \ge \lambda_{\max}^j,
\end{equation*}
which implies $\lambda_{\max} \le (m_j)^{1/j}$. To establish a robust upper bound, we consider the maximum possible value for $m_j$, which is $\hat{m}j + \varepsilon$. This yields:
\begin{equation*}
\lambda{\max} \le (\hat{m}j + \varepsilon)^{1/j}.
\end{equation*}
Since this must hold for all $j \in {2, \dots, k}$, we take the tightest possible bound:
\begin{equation*}
\lambda{\max} \le \min_{2 \le j \le k} (\hat{m}_j + \varepsilon)^{1/j}.
\end{equation*}

The lower bound is constructed from the maximum of two distinct estimates. This composite form provides a tighter bound than either estimate alone.

1. Ratio-Based Bound: We start by relating consecutive moments:
\begin{equation*}
m_{j+1} = \sum_i \lambda_i^{j+1}  \le \lambda_{\max} \sum_i \lambda_i^j = \lambda_{\max} m_j.
\end{equation*}
This gives the relation $\lambda_{\max} \ge m_{j+1}/m_j$. To find the worst-case lower bound, we minimize this ratio by taking the minimum value for the numerator ($m_{j+1} \ge \hat{m}{j+1} - \varepsilon$) and the maximum for the denominator ($m_j \le \hat{m}j + \varepsilon$):
\begin{equation*}
\lambda{\max} \ge \frac{\hat{m}{j+1} - \varepsilon}{\hat{m}_j + \varepsilon}.
\end{equation*}
Taking the maximum over all valid ratios gives the tightest bound from this method.

2. Root-Based Bound: We again start from the moment definition:
\begin{equation*}
m_j = \sum_i \lambda_i^j = \sum_i \lambda_i \cdot \lambda_i^{j-1} \le \lambda_{\max}^{j-1} \sum_i \lambda_i = \lambda_{\max}^{j-1},
\end{equation*}
where we used the trace condition $\mathrm{Tr}(\rho) = \sum_i \lambda_i = 1$. This implies $\lambda_{\max} \ge (m_j)^{1/(j-1)}$. To guarantee a lower bound, we use the minimum value for $m_j$, which is $\hat{m}j - \varepsilon$:
\begin{equation*}
\lambda{\max} \ge (\hat{m}_j - \varepsilon)^{1/(j-1)}.
\end{equation*}
Again, we take the maximum over all $j \in {2, \dots, k}$ to obtain the tightest bound. Combining these two lower bound estimates gives the final expression presented in the main text.

\subsection{Additional Experimental Results for Quantum Virtual Cooling}
\label{fig:qvc_shots}
\begin{figure*}
    \centering
    \includegraphics[width=0.8\linewidth]{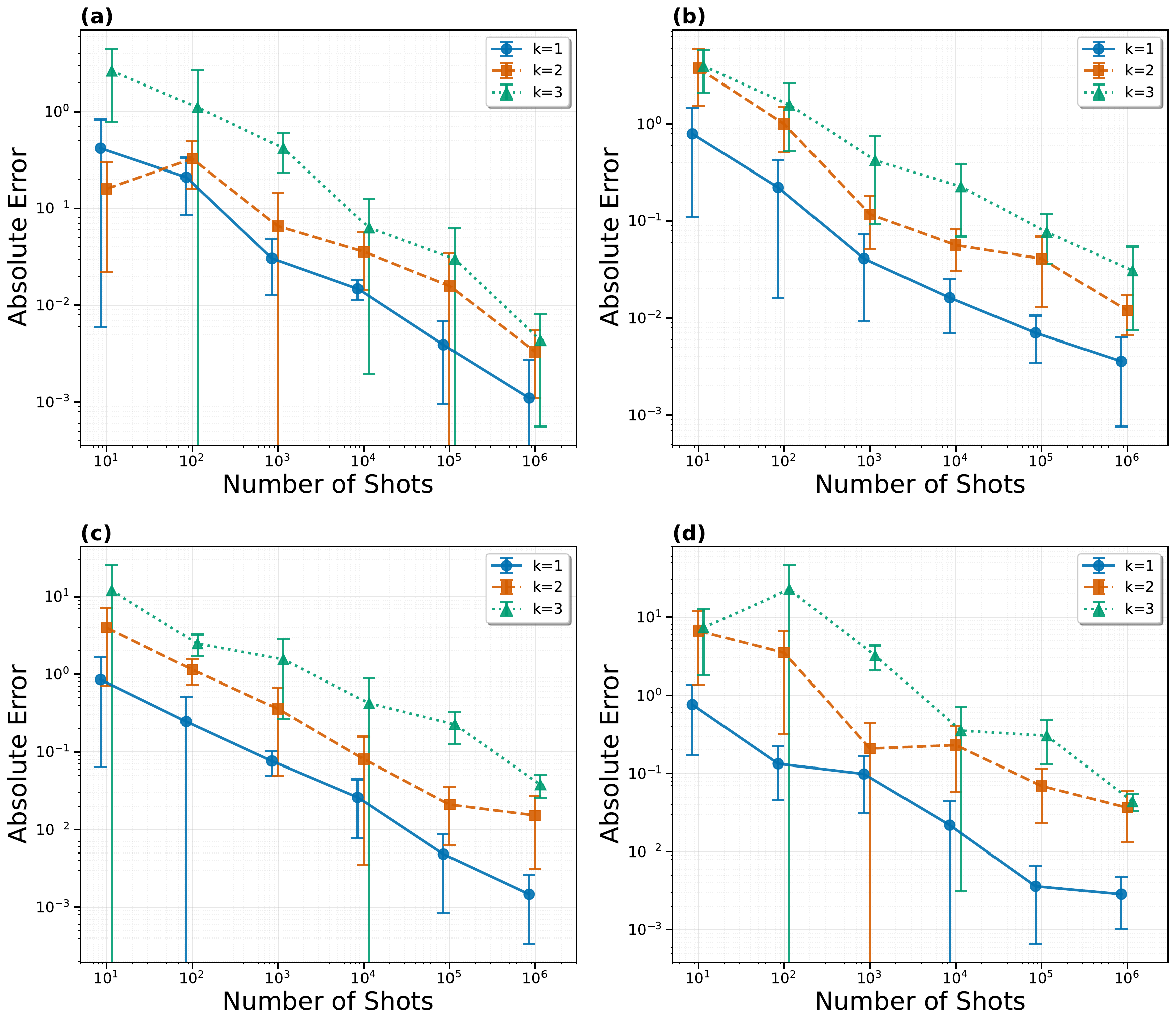}
    \caption{Error scaling analysis of the quantum virtual cooling algorithm for the one-dimensional Heisenberg model. Panels (a)-(d) show the absolute error of the simulated energy expectation value as a functional of the number of measurement shots for system sizes of (a) $n = 3$, (b) $n = 4$, (c) $n = 5$, and (d) $n = 6$ qubits, respectively. All simulations were performed at an inverse temperature $\beta = 0.5$ with coupling parameters $J = h = 1$.}
    \label{fig:appendix_qvc_2}
\end{figure*}

\begin{figure*}
    \centering
    \includegraphics[width=0.9\linewidth]{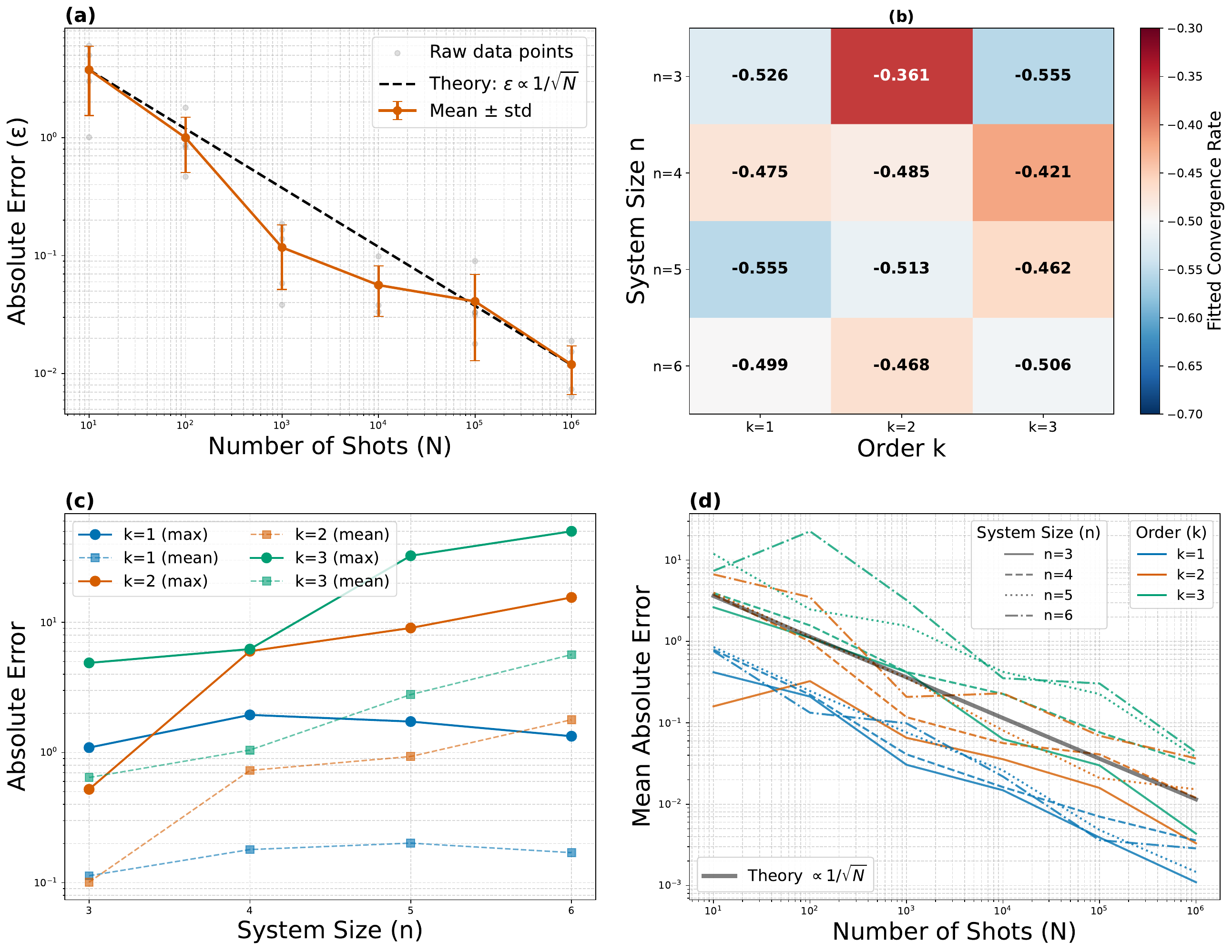}
    \caption{The figure provides comprehensive evidence that the algorithm achieves the standard quantum limit scaling and is scalable with system size. (a) Absolute error $\epsilon$ versus number of shots $N$ for a representative case ($n=4, k=2$). (b) Heatmap of the fitted convergence rate $\alpha$ from the scaling law $\epsilon \propto N^{\alpha}$. (c) Mean and maximum error as a function of system size $n$. (d) Mean error for all configurations, showing a universal collapse onto the $\epsilon \propto N^{-1/2}$ trend (thick reference line), which confirms the generality of the scaling behavior.}
    \label{fig:appendix_qvc}
\end{figure*}

To assess the performance and statistical robustness of our protocol, we conducted numerical simulations using a one-dimensional Heisenberg model ($J=h=1$). A detailed analysis of the results, presented in Figure~\ref{fig:appendix_qvc_2}, helps to illuminate the statistical features of estimating the cooled energy expectation values, $\langle H \rangle_k \equiv \operatorname{Tr}(H\rho^k) / \operatorname{Tr}(\rho^k)$. A key consideration is that $\langle H \rangle_k$ is a ratio of two independently estimated quantities, making its statistical error sensitive to fluctuations in both the numerator and the denominator. This effect is compounded by a physical property of the protocol: as the cooling order $k$ increases, the denominator $\operatorname{Tr}(\rho^k)$ (the purity of the $k$-th power of the state $\rho$) systematically decreases for any mixed state. This smaller denominator can act as an amplifier for statistical noise, potentially leading to significant variance in the final estimate, particularly at low shot counts.

This noise amplification mechanism is quantitatively illustrated across the subplots in Figure~\ref{fig:appendix_qvc_2}. For the $n=4$ system [Figure~\ref{fig:appendix_qvc_2}(b)] with a budget of $N=10^3$ shots, the mean absolute error for $k=1$ is approximately $0.14$, which increases to $0.48$ for $k=2$ and to $2.02$ for $k=3$. This increase underscores the practical challenge that a larger shot budget may be required to achieve a target precision for higher-order estimators. In some resource-limited regimes, the estimator can produce unreliable results; for instance, for the $n=4$ system with $N=10^3$ shots, the absolute error for the $k=3$ estimator reaches 2.02, an order of magnitude larger than the corresponding error of 0.14 for the standard $k=1$ estimator.

Despite this inherent variance at low shot counts, the data in Figure~\ref{fig:appendix_qvc_2} also show the protocol's long-run robustness. As the shot budget increases, the absolute error systematically converges towards zero, even in the more challenging scenarios. Revisiting the $n=5, k=3$ case [Figure~\ref{fig:appendix_qvc_2}(c)], the error drops from $1.57$ at $10^3$ shots to $0.43$ at $10^4$ shots, and finally reaches $0.02$ at $10^6$ shots. This consistent convergence across all tested configurations suggests a universal underlying scaling behavior.

This observation motivated a more general analysis, summarized in Figure~\ref{fig:appendix_qvc}. We find that the convergence of the absolute error $\epsilon$ with the number of measurement shots $N$ consistently follows the standard quantum limit. Figure~\ref{fig:appendix_qvc}(a) shows a representative case where the error closely follows the expected $\epsilon \propto N^{-1/2}$ scaling. To confirm this is a general feature, we fitted the error to a power-law $\epsilon \propto N^{\alpha}$ for all twelve configurations of system size ($n$) and cooling order ($k$). The resulting exponents, shown in the heatmap of Figure~\ref{fig:appendix_qvc}(b), cluster tightly around the theoretical value of $\alpha=-0.5$. Furthermore, Figure~\ref{fig:appendix_qvc}(c) demonstrates that the error has a weak dependence on system size, suggesting the algorithm is scalable. The universality of this behavior is visually captured in Figure~\ref{fig:appendix_qvc}(d), where the mean errors for all configurations collapse onto a single master curve following the $N^{-1/2}$ trend.

In summary, while the estimation of higher-order moments presents practical challenges due to statistical noise amplification at low shot counts, our findings support the protocol's theoretical soundness. The error reliably converges and scales according to the standard quantum limit, allowing for arbitrary precision to be achieved with sufficient measurement resources. This quantitative characterization of the trade-offs is valuable for applying our method to estimate higher-order moments, which are relevant for probing entanglement and non-classical correlations in many-body systems.

\subsection{Experimental Demonstration on a Superconducting Quantum Processor}
\label{appendix_B}
The integer R\'{e}nyi entropy constitutes a fundamental family of information measures that provide crucial insights into quantum states beyond conventional von Neumann entropy. 
For a quantum state $\rho$ and integer order $\alpha \geq 2$, the integer R\'{e}nyi entropy is defined as
\begin{equation*}
S_\alpha(\rho) = \frac{1}{1-\alpha} \log \mathrm{Tr}(\rho^\alpha).
\end{equation*}

\begin{figure*}[htbp]
    \centering
    \begin{tabular}{cc}
        \begin{minipage}{0.45\textwidth}
            \centering
            \includegraphics[width=\linewidth]{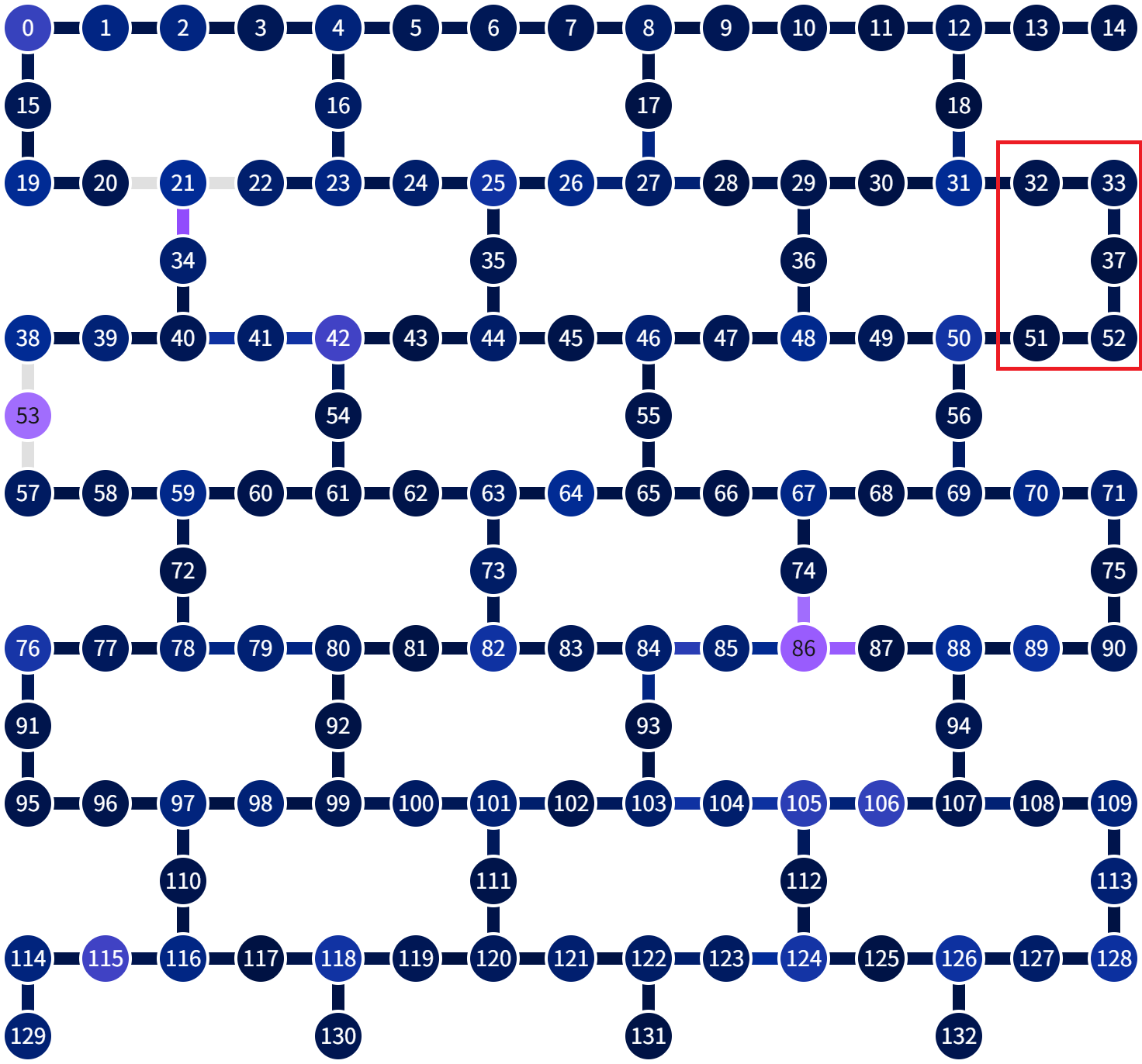}
            (a)
        \end{minipage} &
        \begin{minipage}{0.45\textwidth}
            \centering
            \includegraphics[width=\linewidth]{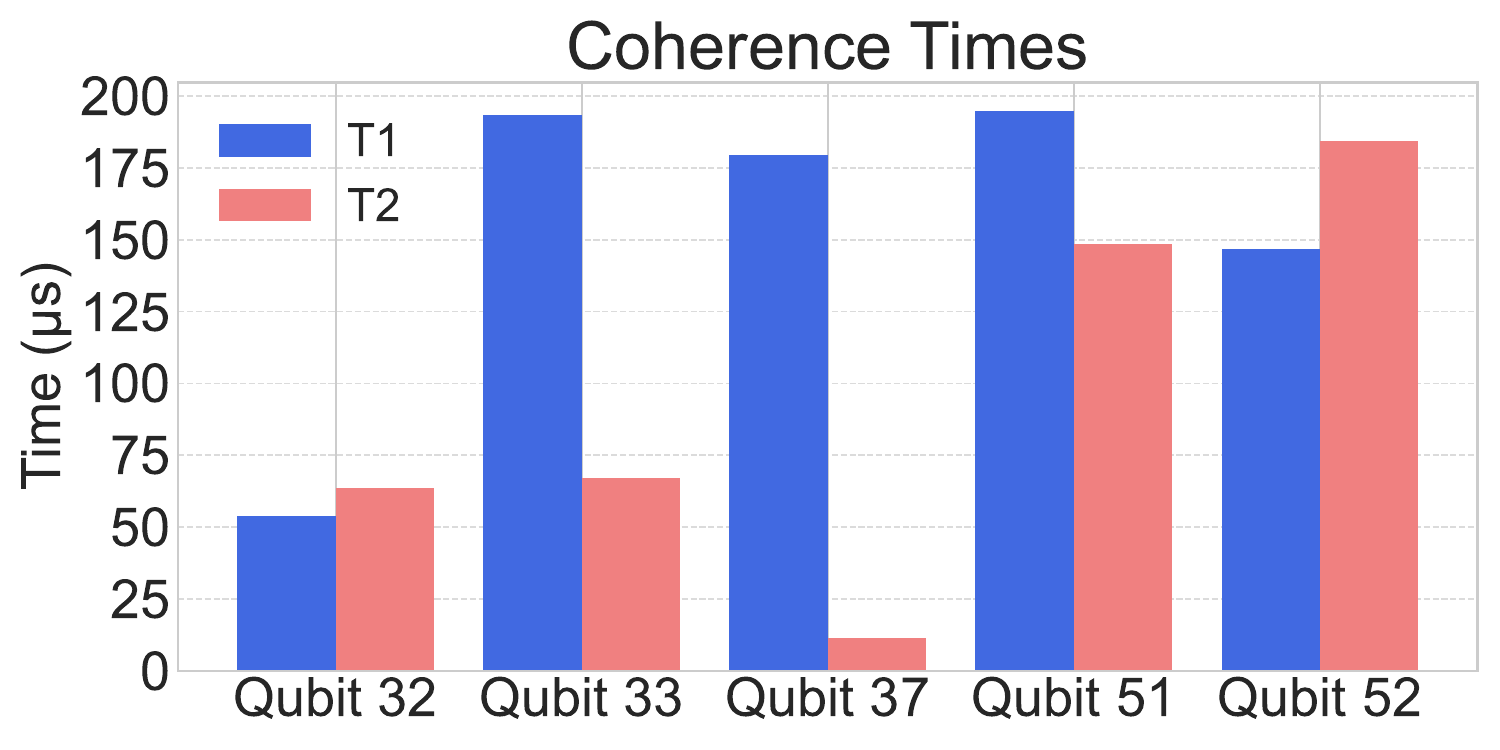}\\
            (b)\\[0.5em]
            \includegraphics[width=\linewidth]{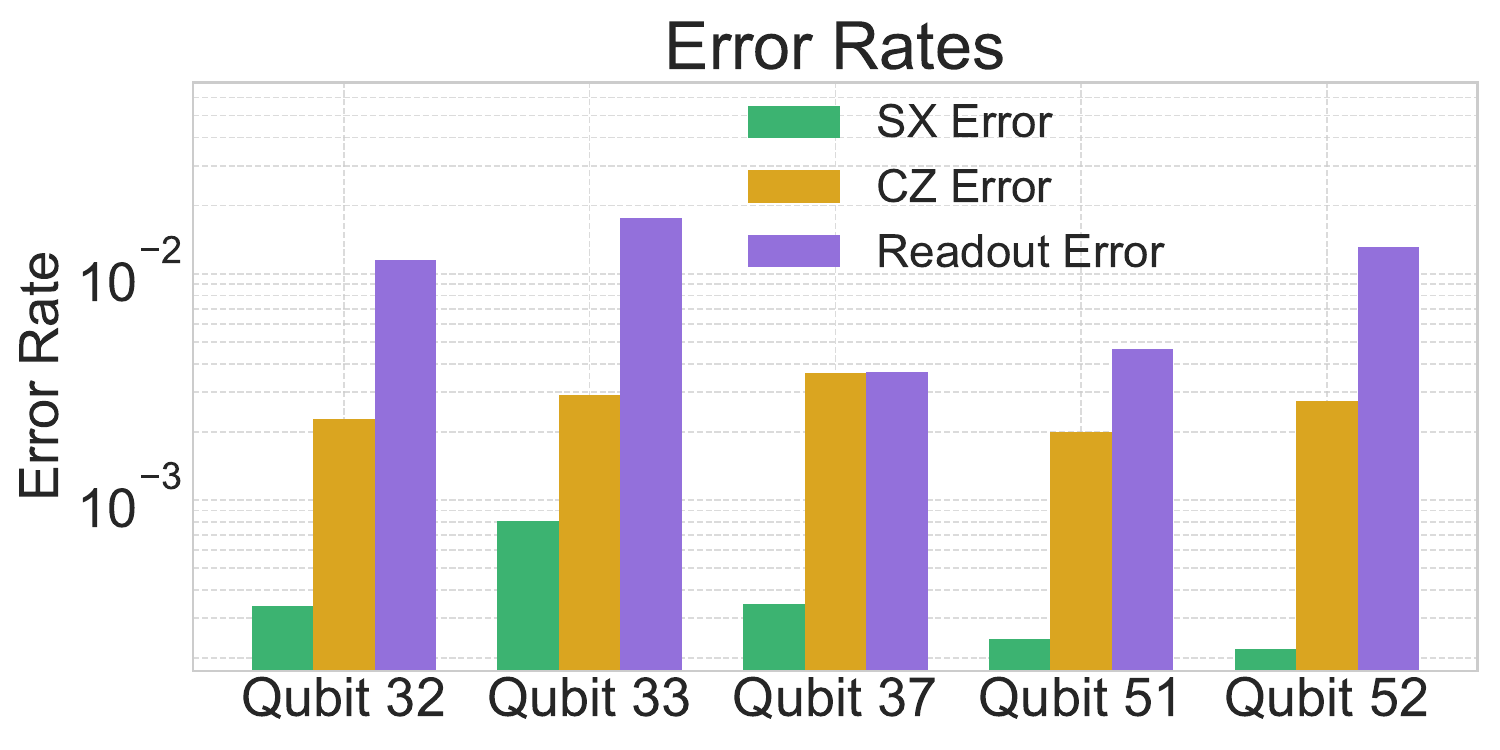}\\
            (c)
        \end{minipage}
    \end{tabular}
    \caption{(a) Physical connectivity map of \texttt{ibm\_torino}; the five qubits employed in this work are boxed in red. 
    (b) Energy-relaxation ($T_1$) and dephasing ($T_2$) times of the selected qubits, averaged over the calibration window used for the experiment.
    (c) Native single-qubit gate (SX), two-qubit controlled-phase (CZ), and simultaneous read-out error rates for the same qubits.
    }
    \label{fig:torino_circuit}
\end{figure*}

\begin{figure*}[htbp]
    \centering
    \includegraphics[width=0.95\columnwidth]{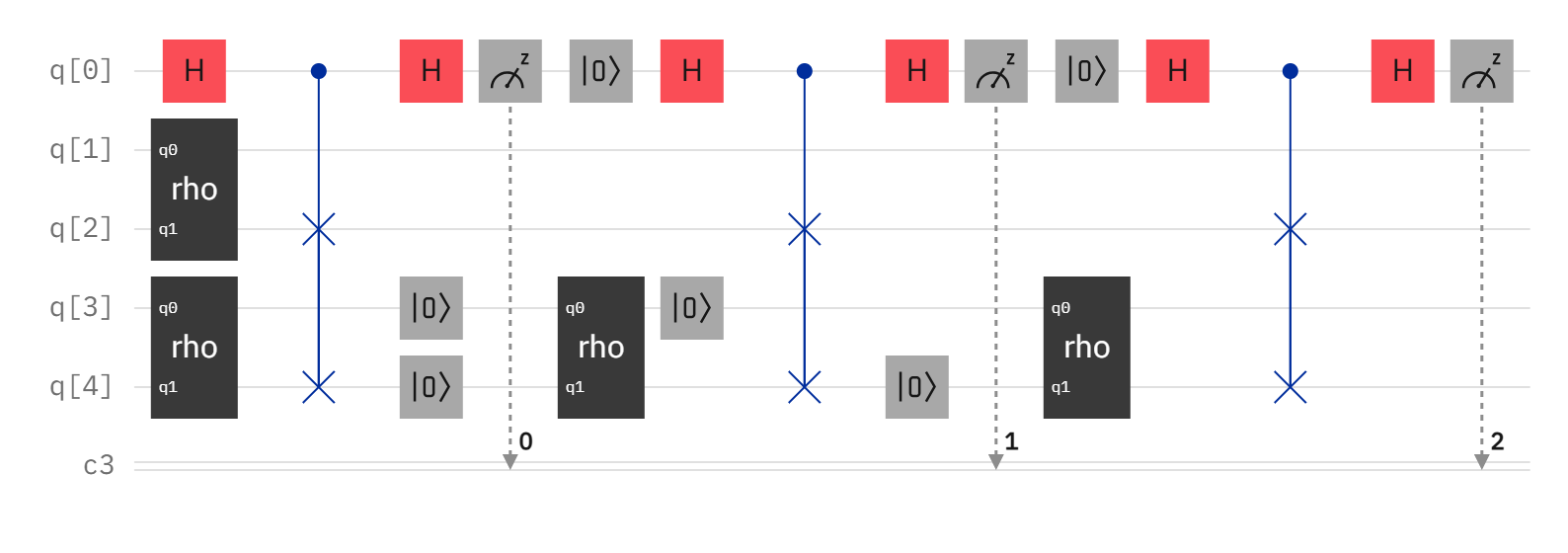}
    \caption{The block $\rho$ prepares the Gibbs state $\rho(\beta)$ with $\beta=0.5$ using an $R_{y}$ rotation followed by a CNOT between the two physical qubits. 
    The subsequent CSWAP realizes the permutation operator; the number of ancilla-controlled layers grows linearly with the R\'{e}nyi order $\alpha$. 
    $Z$-basis measurements on the ancilla supply the bit-strings used to compute $\operatorname{Tr}(\rho^{\alpha})$.
    }
    \label{fig:circuit_exp}
\end{figure*}

We perform this experiment on \verb|ibm_torino|~\cite{ibm_torino}, a $133$-qubit superconducting quantum processor.
Details can be found in Fig.~\ref{fig:torino_circuit}.
The circuit we use is shown in Fig.~\ref{fig:circuit_exp}.
The state we chose is the Gibbs state for $H = Z$.
That is,
\[
\rho = \begin{bmatrix}
    \mathrm{e}^{-\beta} /( \mathrm{e}^{-\beta} +\mathrm{e}^{\beta}) &\\
    & \mathrm{e}^{\beta} /( \mathrm{e}^{-\beta} +\mathrm{e}^{\beta})
\end{bmatrix}.
\]

\begin{figure*}[htbp]
    \centering
    \begin{tabular}{cc}
         \includegraphics[width=0.45\columnwidth]{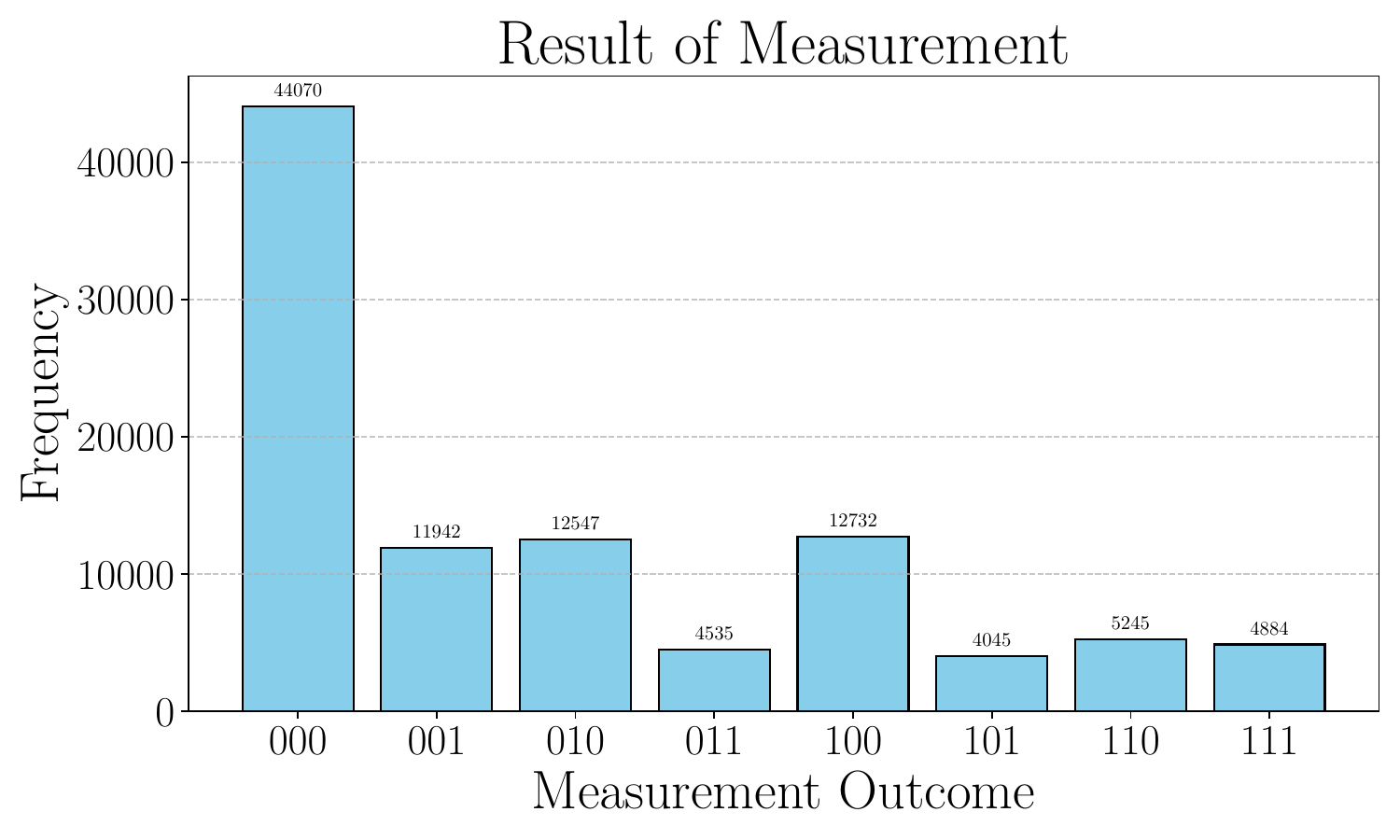}& \includegraphics[width=0.45\columnwidth]{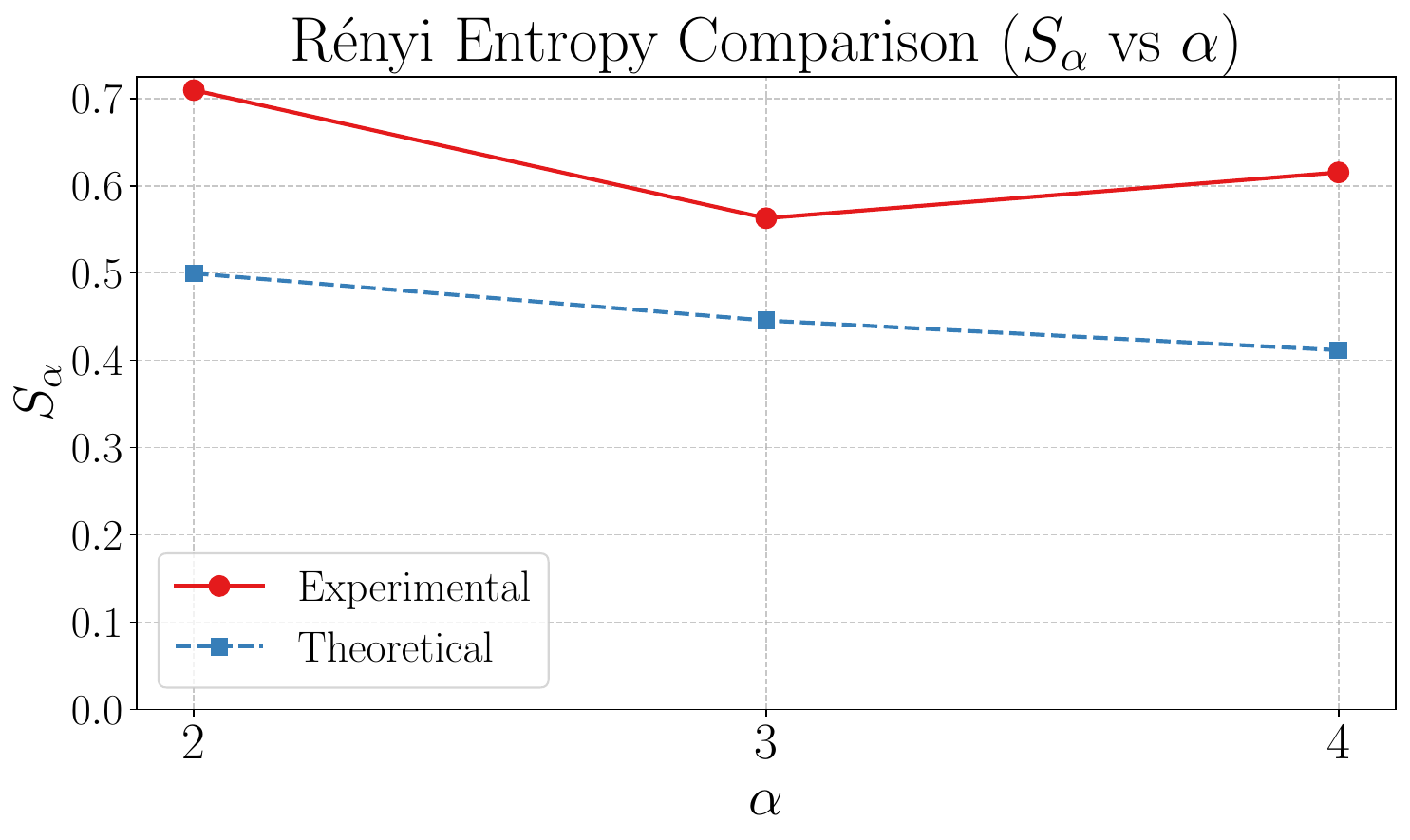} \\
        (a) & (b) 
    \end{tabular}
    \cprotect\caption{Output statistics of the Gibbs-state integer R\'{e}nyi entropy.
    (a) Histogram of measurement outcomes obtained from $10^5$ shots of the circuit on the \verb|ibm_torino| processor. 
    (b) Comparison between the exact R\'{e}nyi entropy and the values extracted experimentally integer orders $\alpha=2,3,4$.}
    \label{fig:result_exp_ibm}
\end{figure*}

To prepare this state on two qubits, we only need to perform an $R_y(\theta)$ gate on the first qubit and a $CX$ gate to entangle the two qubits, where $\theta = 2\arctan \mathrm{e}^{\beta}$.

Taking $\beta = 0.5$, we ran the circuit $1$ million times and the result can be seen in Fig.~\ref{fig:result_exp_ibm}.
The demonstration experiment was performed on July $24^{\mathrm{th}}$, $2025$.

The experimental result of $\mathrm{Tr}(\rho^3)$ is $0.32442$ while the theoretical value is $0.41016$.
Consider a state $\sigma = \mathrm{diag}[0.6575, 0.3425]$, we have $\mathrm{Tr}(\sigma^3)\approx 0.3244$.
The fidelity between $\rho$ and $\sigma$ is roughly $0.85$.
Although the fidelity is still considerable, the result is not that good.
This explains why there is a gap between the experimental result and the theoretical value.
Furthermore, as $\alpha$ increases, the experimental values rise due to noise, while the theoretical values decrease.

\end{document}